\documentclass[floats,floatfix,amssymb,prd,twocolumn,nofootinbib,superscriptaddress]{revtex4-1}
\usepackage{amssymb,amsmath,verbatim,mathtools,needspace,enumitem,etoolbox,graphicx,physics,microtype,ulem}
\usepackage{xspace}
\usepackage{multirow}
\usepackage[dvipsnames, usenames]{xcolor}
\usepackage{xspace}
\definecolor{linkcolor}{rgb}{0.0,0.3,0.5}
\definecolor{urlcolor}{rgb}{0.27,0.55,0.}
\definecolor{funcolor}{rgb}{0.65, 0.16, 0.16}

\usepackage[unicode, colorlinks=true, linkcolor=linkcolor, citecolor=linkcolor, filecolor=linkcolor,urlcolor=linkcolor, pdfusetitle]{hyperref}
\usepackage[all]{hypcap}
\usepackage{aas_macros}
\usepackage{subcaption}
\usepackage[utf8]{inputenc}
\usepackage{lineno}
\usepackage{natbib,tabularx}
\usepackage{booktabs}%

\newcolumntype{b}{>{\hsize=0.6\hsize\centering\arraybackslash}X}

\newcommand{\chieff}{\ensuremath{\chi_{\mathrm{eff}}}\xspace}

\newcommand{\msun}{\ensuremath{M_{\odot}}\xspace}
\newcommand{\mc}{\ensuremath{\mathcal{M}}\xspace}

\newcommand{\beqn}{\begin{eqnarray}}
\newcommand{\enqn}{\end{eqnarray}}
\newcommand{\beq}{\begin{equation}}
\newcommand{\eeq}{\end{equation}}

\newcommand{\linf}{\textsc{LALInference}}
\newcommand{\lk}{\ensuremath{\mathcal{L}}}

\newcommand{\LIGOlabMIT}{\affiliation{LIGO Laboratory, Massachusetts Institute of Technology, 185 Albany St, Cambridge, MA 02139, USA}}
\newcommand{\MKI}{\affiliation{Department of Physics and Kavli Institute for Astrophysics and Space Research, Massachusetts Institute of Technology, 77 Massachusetts Ave, Cambridge, MA 02139, USA}}
\newcommand{\Austin}{\affiliation{Theory Group, Department of Physics, University of Texas at Austin, Austin, TX 78712, USA}}
\newcommand{\IASaffiliation}{\affiliation{School of Natural Sciences, Institute for Advanced Study, 1 Einstein Drive, Princeton, NJ 08540, USA}}
\newcommand{\Princeton}{\affiliation{Department of Physics, Princeton University, Princeton, NJ, 08540, USA}}

\usepackage{gensymb} %
\usepackage{amsbsy} %

\begin{document}
\title{Source properties of the lowest signal-to-noise-ratio binary black hole detections} 

\author{Yiwen Huang}
\email{ywh@mit.edu}
\LIGOlabMIT \MKI

\author{Carl-Johan Haster}
\email{haster@mit.edu}
\LIGOlabMIT \MKI

\author{Salvatore Vitale}
\email{salvatore.vitale@ligo.org}
\LIGOlabMIT \MKI

\author{Aaron Zimmerman}
\email{aaron.zimmerman@austin.utexas.edu}
\Austin

\author{Javier Roulet}
\Princeton
\author{Tejaswi Venumadhav}
\IASaffiliation
\author{Barak Zackay}
\IASaffiliation
\author{Liang Dai}
\IASaffiliation
\author{Matias Zaldarriaga}
\IASaffiliation

\date{\today}

\begin{abstract}

We perform a detailed parameter estimation study of binary black hole merger events reported in \citet{Zackay:2019tzo} and \citet{Venumadhav:2019lyq}. 
These are some of the faintest signals reported so far, and hence, relative to the loud events in the GWTC-1 catalog~\cite{LIGOScientific:2018mvr}, the data should have lesser constraining power on their intrinsic parameters. 
Hence we examine the robustness of parameter inference to choices made in the analysis, as well as any potential systematics. 
We check the impact of different methods of estimating the noise power spectral density, different waveform models, and different priors for the compact object spins.
For most of the events, the resulting differences in the inferred values of the parameters are much smaller than their statistical uncertainties. 
The estimation of the effective spin parameter \chieff, i.e. the projection of the mass-weighted total spin along the angular momentum, can be sensitive to analysis choices for two of the sources with the largest effective spin magnitudes, GW151216 and GW170403.
The primary differences arise from using a 3D isotropic spin prior: the tails of the posterior distributions should be interpreted with care and due consideration of the other data analysis choices.
\\

\end{abstract}

\maketitle

\section{Introduction}

The LIGO~\cite{TheLIGOScientific:2014jea} and Virgo~\cite{TheVirgo:2014hva} collaborations (LVC) have to date reported the detection of 10 binary black hole (BBH) systems and one binary neutron star (BNS) system in the data collected during their first two observing runs~\cite{LIGOScientific:2018mvr}.
An independent analysis of the public data released by the LVC~\cite{Vallisneri:2014vxa,O1dataRelease,O2dataRelease,Abbott:2019ebz} has revealed 9 additional potential gravitational-wave (GW) signals~\cite{Zackay:2019tzo,Venumadhav:2019lyq,Zackay:2019btq} (see also Ref.~\cite{Nitz:2018imz} for a re-analysis of LIGO--Virgo's first observing run and Ref.~\cite{Nitz:2019hdf} for the second observing run).
In this study we focus on the 7 signals presented in \citet{Zackay:2019tzo} and \citet{Venumadhav:2019lyq}.
If of astrophysical origin, then these systems are also BBHs, thus nearly doubling the total number of BBH systems detected in the first two observing runs.

The parameters of observed BBHs~\cite{Venumadhav:2019lyq,LIGOScientific:2018jsj} encode information about the underlying BBH population and about the evolutionary history of the black holes and their progenitors. 
The masses and spins of the black holes in particular can be used to infer the formation mechanism of the observed binaries.
Usually, two families of formation scenarios are considered: classical binary evolution in the galactic field
~\cite{Nelemans:2001hp,Belczynski:2001uc,Voss:2003ep,Belczynski:2006zi,Belczynski:2005mr,Dominik:2013tma,Belczynski:2014iua,Mennekens:2013dja,Spera:2015vkd,Eldridge:2016ymr,Stevenson:2017tfq,Mapelli:2017hqk,Giacobbo:2017qhh,Giacobbo:2018etu,Kruckow:2018slo}, or dynamical formation either in the galactic field~\cite{Silsbee:2016djf}, or in dense environments such as clusters~\cite{Barack:2018yly,Varri2018ModestReview} or AGN disks~\cite{Antonini:2012ad,McKernan:2012rf,Stone:2016wzz,Bartos:2016dgn}.\footnote{Other possibilities exist: e.g. primordial black holes~\cite{Barack:2018yly,Varri2018ModestReview}.}
This latter scenario could also result in repeated mergers, which would produce heavier black holes~\cite{Chatziioannou:2019dsz,Kimball:2019mfs,Rodriguez:2019huv}. 

The spins of the black holes, and specifically the relative orientation of black hole spins in a binary, can be used to discriminate between these formation channels: formation in the field is expected to result in spins which are nearly aligned with the orbital angular momentum (if tides are efficient in spinning up the progenitors), while dynamical formation should not set any such preferential direction~\cite{Rodriguez:2016vmx}. 
Unfortunately, it is often hard to measure the \emph{individual} spins of black holes in binaries with any significant precision~\cite{Vitale:2014mka,Purrer:2015nkh,Vitale:2016avz}. 
While it is still possible to measure the relative occurrence of BBHs in the different formation channels using the component spins and their orientation, hundreds of detections would be required before a firm measurement can be achieved~\cite{Vitale:2015tea,Talbot:2017yur,Stevenson:2017dlk}.

While individual black hole spins ($\boldsymbol{S}$) are difficult to measure, a combination of the two spins called the effective spin \chieff~\cite{Damour:2001tu,Racine:2008qv,Santamaria:2010yb,Ajith:2009bn} is usually much better measured~\cite{Vitale:2016avz,LIGOScientific:2018mvr,LIGOScientific:2018jsj,Ng:2018neg,Zackay:2019tzo,Venumadhav:2019lyq}.
The effective spin is the mass-weighted projection of the dimensionless spins of the components, $\boldsymbol {\chi}_i = c \boldsymbol{S}_i/G$, along the orbital angular momentum $\boldsymbol{L}$:
\begin{equation}
\chieff= \left(\frac{m_1 \boldsymbol{\chi}_1+ m_2 \boldsymbol{\chi}_2}{m_1+m_2}\right)\cdot\frac{\boldsymbol{L}}{|\boldsymbol{L}|} \,.
\end{equation}
Formation channels that preferentially align the spins with the orbital angular momentum should thus have positive values of \chieff. 
This is not necessarily true for dynamically formed BBHs: since for those all black hole spins orientations are equally likely, the expected distribution for \chieff is centered around zero.
The effective spin can thus be used to infer the astrophysical origin of individual sources, and to reconstruct the overall population of black holes in binaries and of their progenitors~\cite{Farr:2017gtv,Farr:2017uvj,LIGOScientific:2018jsj}.

Remarkably, all of the BBHs reported by the LVC to date are consistent with having small or zero \chieff~\cite{LIGOScientific:2018mvr} at 90\% confidence. 
The two sources for which the largest \chieff was measured are GW151226 ($0.2^{+0.2}_{-0.1}$, median and 90\% credible interval) and GW170729 ($0.4^{+0.2}_{-0.3})$. 
Among the BBHs reported by Ref.~\cite{Zackay:2019tzo,Venumadhav:2019lyq, Zackay:2019btq}, four signals have appreciable \chieff: GW151216, GW170403, GW170121 and GW170817B. 
Especially remarkable are the spins reported for GW151216 and GW170403, where GW151216 was reported as having a large and positive effective spin of $\chieff = 0.8^{+0.1}_{-0.2}$~\cite{Zackay:2019tzo}, while $\chieff = -0.7^{+0.5}_{-0.3}$~\cite{Venumadhav:2019lyq} was inferred for GW170403, making it the largest negative effective spin BBH so far.

The algorithms in Refs.~\cite{Venumadhav:2019tad, Venumadhav:2019lyq} were optimized to detect the faintest individually observable events in the population of merging BBHs, and hence several of the detected signals had relatively modest values of the signal-to-noise ratio (SNR).
As the information content of observed signals scales with their  SNR\footnote{Squared information, as defined in the sense of Shannon's information theory, is proportional to the squared SNR.},
the data will have the least constraining power on the intrinsic parameters for faint events such as GW151216 and GW170403.
In light of this fact, it is worth carefully considering the impact on the inferred parameters of the various analysis choices adopted in parameter estimation: the Bayesian priors, GW waveform model, and the treatment of the instrumental noise, in particular its power spectral density (PSD)~\cite{Sathyaprakash:2009xs}.
For example, Ref~\cite{Vitale:2017cfs} has shown how the \chieff measurement of the LVC detection GW151012~\cite{LIGOScientific:2018mvr} is sensitive to the prior choice; while Ref.~\cite{Chatziioannou:2019zvs} has shown how the treatment of the noise PSD can impact the source characterization analysis. 

In this paper we analyze all of the BBHs reported by Refs.~\cite{Zackay:2019tzo,Venumadhav:2019lyq}. 
We perform parameter estimation with the procedures discussed in~\cite{Zackay:2019tzo,Venumadhav:2019lyq,RelativeBinning}, as well as the ones used by the LVC in the analysis of GWTC-1~\cite{LIGOScientific:2018mvr}. %
We use three distinctive sets of analysis choices including method of PSD estimation, sampling algorithms, signal models, and prior assumptions for the source parameters.

Our results underline that the specific configuration of the analysis can have a significant impact on the astrophysical inference of some of the BBHs detected to date, especially if they have low SNRs.
In particular, we show that the high \chieff of GW151216 and GW170403 can be significantly reduced depending on the spin priors used in the analysis. %
The tails of the distribution need to be interpreted with care and in the context of the analysis choices made, such as the method used to estimate the PSD and the length of data analyzed.
For studies that build on the estimated parameter distributions for such sources, it is important to be aware of these analysis choices before interpreting the results.

\section{Method}
\label{Method}

We treat the detectors' data $d$ as composed of $h$, the putative GW signal, and $n$, the noise,
\begin{equation}
d = h + n \,.
\end{equation}
The GW signal emitted by a compact binary in a quasi-circular orbit can be described by a model waveform, with 15 parameters\footnote{
We do not consider tidal deformability of the compact objects, since in this work we assume all binary sources to contain only black holes.},
including masses, spins, sky position, luminosity distance and orbital orientation. 
We use Bayesian inference to measure the parameters of signals embedded in the data~\cite{Veitch:2014wba,TheLIGOScientific:2016wfe}.

The end result of parameter estimation is a posterior probability density function (PDF) for the unknown parameters $\boldsymbol {\theta}$:

\begin{equation}\label{Eq.BayesTheorem}
p({\boldsymbol {\theta}}|d, H) = \frac{\pi({\boldsymbol {\theta}}|H)\lk(d|{\boldsymbol {\theta}},H)}{Z(d | H)}
\end{equation}
where $\pi({\boldsymbol {\theta}}|H)$ is the prior probability density of ${\boldsymbol {\theta}}$ given the hypothesis $H$, $Z(d | H)$ is a normalization factor, and $\lk(d |{\boldsymbol {\theta}},H)$ is the likelihood that we take as describing stationary and Gaussian noise

\begin{equation}\label{Eq.Likelihood}
\lk({\boldsymbol {d}}|{\boldsymbol {\theta}},H) \propto \exp(-\frac{1}{2} \langle d - h({\boldsymbol\theta})| { d - h({\boldsymbol\theta})} \rangle),
\end{equation}
where the quantity in angle brackets represents a noise weighted inner product
\begin{equation}\label{Eq.Inner Product}
\langle a(\boldsymbol {\theta}, f)|b(\boldsymbol {\theta}, f) \rangle \equiv 2\int^{f_\mathrm{high}}_{f_\mathrm{low}}\frac{a(\boldsymbol {\theta},f) b(\boldsymbol {\theta},f)^*+ {\rm c.c.}}{S_n(f)} \mathrm{d}f .
\end{equation}
The PSD of the detector noise, here labeled as ${S_n(f)}$, is the Fourier transformed autocorrelation of the time-domain detector noise~\cite{Sathyaprakash:2009xs}, and must be estimated from the data (see below for more details).

The evidence of the data, $Z$, is the normalization constant in Eq.~\eqref{Eq.BayesTheorem},
\begin{equation}
Z(d|H) = \int d\theta_1...d\theta_N \, p(d|{\boldsymbol {\theta}},H) p(\boldsymbol{\theta}|H).
\label{Eq.Evidence}
\end{equation}
Given two alternative models (e.g.~a waveform family that accounts for spin-induced orbital precession, and one that does not), the ratio of their evidences, known as the Bayes factor, can be used to quantify the relative confidence between different models.

The signal-to-noise ratio (SNR) reported in this work is\footnote{Note that SNRs reported by LVC~\cite{LIGOScientific:2018mvr} is defined as {$\frac{\sum_{IFO} \langle d | h \rangle}{\sqrt{\sum_{IFO} \langle h | h \rangle}}$}  instead.}
\begin{equation}
\rho = \sqrt{\sum_{IFO} (2 \langle d | h \rangle - \langle h | h \rangle) }.
\label{Eq.SNR}
\end{equation}
When we have a network with multiple interferometers (IFO), we report the values of network SNR, which we obtain by adding in quadrature the values from each detector.

\begin{table*}[t!]
\centering
\begin{tabularx}{0.95\linewidth}{cbcbc}
\toprule
\toprule
  Configuration & PSD & Sampler & Prior  & Waveform \\ [0.5ex]
 \midrule
   A & \textsc{BayesWave} &  \multirow{2}{*}{\textsc{LALInferencenest}~\cite{Veitch:2009hd,Veitch:2014wba}} & 3D Isotropic spin  &  IMRPhenomPv2~\cite{Husa:2015iqa,Khan:2015jqa,Hannam:2013oca} \\
  B & \cite{Littenberg:2014oda,Cornish:2014kda,Chatziioannou:2019zvs} & & Aligned-spin, $\chi_{z}$ same as Config. A & SEOBNRv4\_ROM~\cite{Bohe:2016gbl} \\
   C & Welch's method w/ drift factor~\cite{Venumadhav:2019tad} & \textsc{pyMultiNest} ~\cite{Feroz:2007kg,Feroz:2008xx,Feroz:2013hea,Buchner:2014nha} & Aligned-spin, flat in \chieff~\cite{Zackay:2019tzo} &  IMRPhenomD~\cite{Husa:2015iqa,Khan:2015jqa} \\
\bottomrule
\bottomrule
\end{tabularx}
\caption{Main differences between the configurations used in this work. More details are provided in the text.}
\label{Table.Configuration}
\end{table*} 

The results we present below are obtained using three different configurations, as detailed in Table \ref{Table.Configuration}, with different choices of sampler, prior, PSD estimation method, and waveform models.
Configs. A and B follow standard procedures of analysis adopted by the LVC in its publications so far. 
In these configurations, we perform the matched-filter analysis using the \textsc{LALSuite} software package~\cite{lalsuite}, and explore the parameter space stochastically using the nested sampling algorithm implemented in \linf~\cite{Veitch:2009hd,Veitch:2014wba}.
The PSDs are {estimated} using the \textsc{BayesWave} algorithm~\cite{Littenberg:2014oda, Cornish:2014kda,Chatziioannou:2019zvs}, over 4 second segments centered around the merger time of each candidate event.
We use data from the same segments to evaluate the likelihood, Eq.~\eqref{Eq.Likelihood}, restricting the domain of the integrals in Eq.~\eqref{Eq.Inner Product} to the frequency range $[20,512]$\,Hz.

In Config. C, we use a matched-filtering approach to perform the parameter estimation (akin to Configs. A and B), but we estimate the PSD over a 4096 second segment of data using Welch's method, with median averaging \cite{Allen:2005fk, Zackay:2019tzo,Venumadhav:2019lyq}.
The exact data used are set by the start and end times of the 4096 seconds long file released by the LVC\footnote{Data made available on \href{https://www.gw-openscience.org/data/}{www.gw-openscience.org}~\cite{Vallisneri:2014vxa,O1dataRelease,O2dataRelease,Abbott:2019ebz}.} that contained the specific event.
Since the statistical properties of the noise typically vary over shorter timescales, in Config. C, we scale the estimated PSD by a local and time-dependent scalar correction (drift factor) when computing the likelihood~\cite{Venumadhav:2019tad, PSDDriftCorrection}. 
We make two other different choices in Config. C as compared to those in A and B: (1) after obtaining the PSD and drift correction factor using the data segment specified above, we compute the likelihood using 128 seconds of data, starting 60\,s before the merger time (restricting to the frequency range $[20,512]$\,Hz, as before), and (2) we explore the space of parameters using \textsc{PyMultiNest}, a standard Python implementation of nested sampling \cite{Feroz:2007kg,Feroz:2008xx,Feroz:2013hea,Buchner:2014nha}.

The configurations also use different waveform models, all of which describe the complete inspiral-merger-ringdown (IMR) of a compact-object binary coalescence, and are calibrated against Numerical Relativity simulations of BBH mergers.
Config. C uses the phenomenological waveform model IMRPhenomD~\cite{Husa:2015iqa,Khan:2015jqa} which assumes BH spins (anti-)aligned to the orbital angular momentum.
The model used by Config. A, IMRPhenomPv2, is constructed from the same aligned-spin model, but is extended to also include an effective description of the effects from spin-precession through a rotation of the underlying IMRPhenomD model~\cite{Husa:2015iqa,Khan:2015jqa,Hannam:2013oca}.
Config. B uses a separately developed aligned-spin only model, SEOBNRv4\_ROM, based on the effective-one-body framework~\cite{Bohe:2016gbl}.

Because the inner product, Eq.~\ref{Eq.Inner Product}, depends on the noise PSD and the waveform model, both of these factors can impact the measured SNR. Keeping everything else the same, we would expect a precessing waveform template to be able to recover more SNR than a spin-aligned one, due to the extra degrees of freedom. This is indeed what we observe comparing Configs. A and B, Sec.~\ref{Results}.
Configs B and C instead use spin-aligned waveforms, but different PSDs and analysis software. In particular, the algorithms we use to estimate the PSD implement different strategies to limit the effect of noise non-stationary and non-Gaussianity~\cite{Venumadhav:2019tad, PSDDriftCorrection,Littenberg:2014oda, Cornish:2014kda,Chatziioannou:2019zvs}. We find that the matched filter SNRs for configs B and C calculated at the maximum likelihood point usually differ by a few percent in either direction.

\begin{figure}[t]
  \centering
    \includegraphics[width=0.49\textwidth]{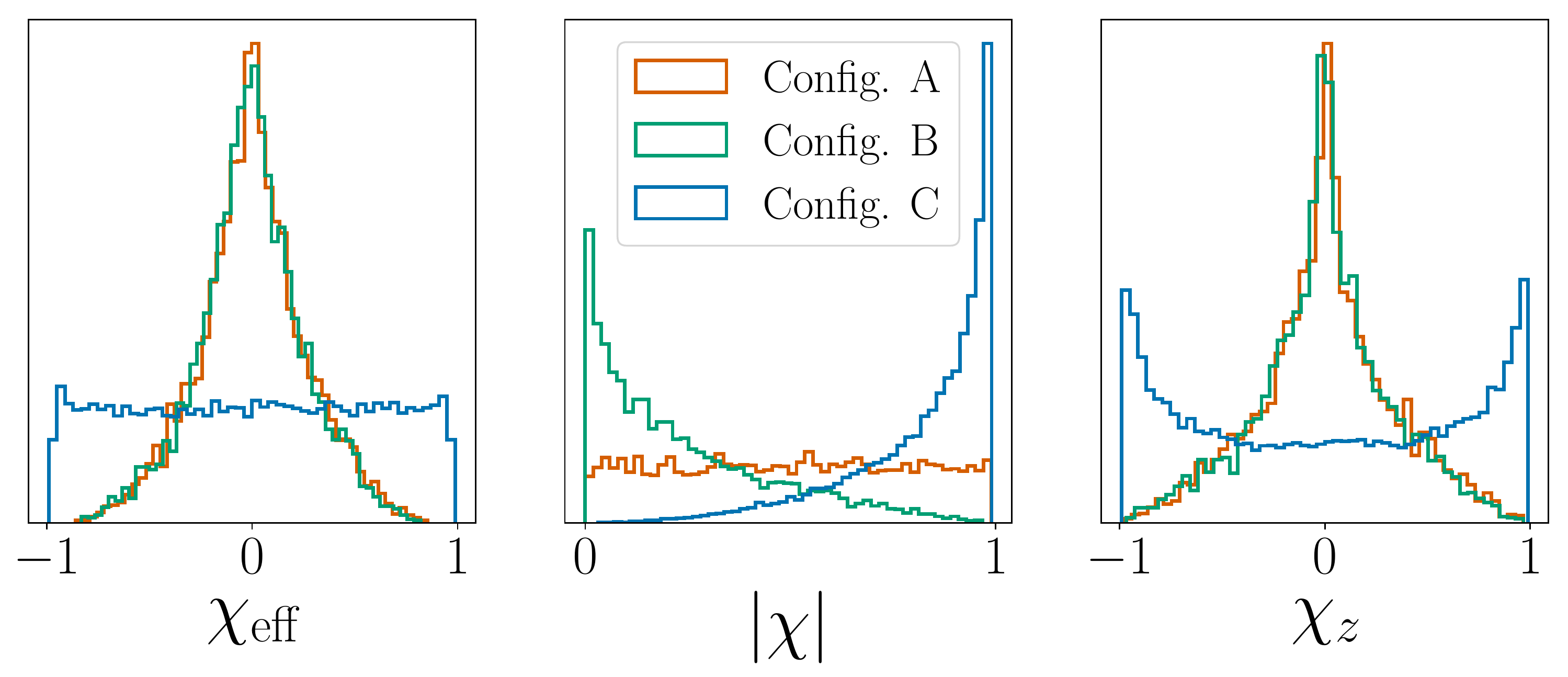}
    \caption{Spin priors on effective spin $\chi_\mathrm{eff}$, individual spin magnitude $|\chi|$, and the $z$-component of spin $\chi_{z}$, used in 3 configurations. 
    Note that for aligned-spin waveforms (Config. B \& C), $|\chi_i| = |\chi_{i,z}| $ where $i = 1,2$ corresponds to individual component of the binary.}
  \label{Fig.spin prior}
\end{figure}

Configs. A and B use priors routinely used in LVC publications~\cite{Abbott:2016blz, TheLIGOScientific:2016wfe, Abbott:2016nmj, Abbott:2017vtc, Abbott:2017gyy, Abbott:2017oio,TheLIGOScientific:2017qsa,Abbott:2018wiz}. 
Config. A (precessing analysis) uses a uniform prior in the dimensionless spin magnitude for each black hole, in the range $[0,0.99]$, and an isotropic prior for the spin orientation. 
Config. B uses a waveform model which assumes aligned spins, and for this case we use the prior from Config. A for the component of the spin along the orbital angular momentum, $\chi_{iz}$.
Finally, Config. C, spin-aligned analysis with IMRPhenomD, uses a spin prior which is uniform in the effective inspiral spin, \chieff.
These prior choices are shown in Fig.~\ref{Fig.spin prior}, where we plot the prior distributions for \chieff, the magnitude of component spins $|\chi|$, and the projection of the primary's spin along the angular momentum ($\chi_{z}$). 
Note that for the two spin-aligned analyses, $|\chi| = |\chi_{z}|$ by definition.

We stress that the prior on \chieff for Config. C is quite different from that for Config. A or B, especially toward the edges. 
Configs. A and B penalize {\it a priori} large \chieff, that is systems for which spins are large in magnitude and nearly aligned with the orbital momentum. 
Conversely, Config. C achieves a prior that is flat in \chieff, by {\it a priori} preferring large individual spin magnitudes.

The three analyses all use similar priors in the other parameters: in particular they all use priors which are uniform over the detector-frame component masses, in a range large enough that the posteriors are not truncated; 
uniform over the sphere for sky position and orientation of the orbit with respect to the line of sight; proportional to the square of the luminosity distance; and uniform in geocenter arrival time and phase.

\section{Results}
\label{Results} 

In this section we report the results of our analyses on all of the gravitational-wave events identified in \citet{Zackay:2019tzo} and \citet{Venumadhav:2019lyq}. 
For all events, we report medians and 90\% credible intervals on the detector frame chirp mass $\mc$, mass ratio $q = m_2/m_1 \in [0,1]$, effective spin \chieff, and luminosity distance $D_L$.
For all configurations, we calculate the network SNRs using Eq.~\eqref{Eq.SNR} and report the values corresponding to the maximum likelihood.
For Configs. A and B, we also report the \emph{natural log} Bayes factor for the gravitational-wave signal model over the Gaussian noise model ($\mathrm{ln}\mathcal{B}_{S/N}$)~\cite{TheLIGOScientific:2016wfe,Veitch:2009hd,DelPozzo:2014cla}. 

\begin{figure}[h]
  \centering
    \includegraphics[width=0.45\textwidth]{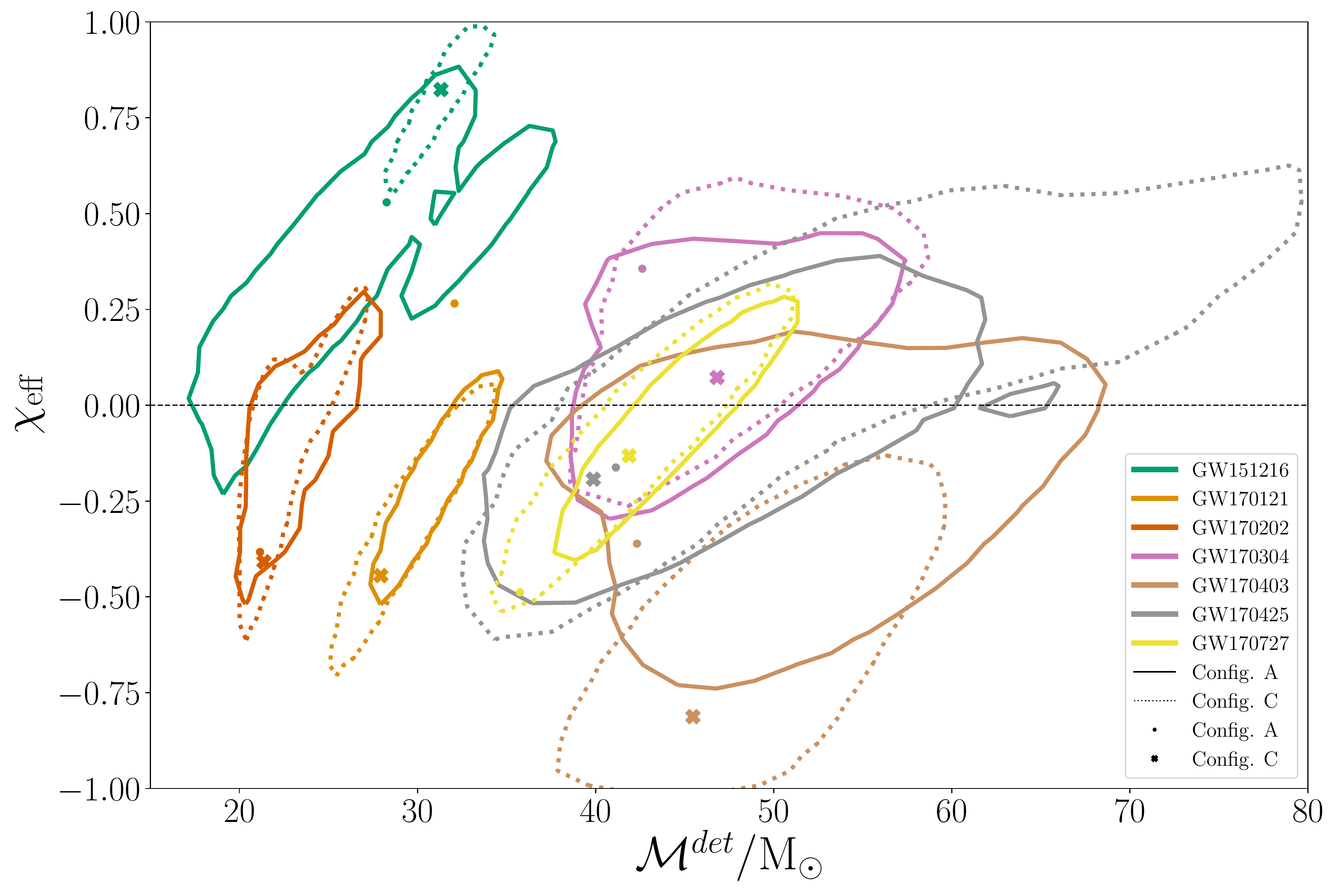}
    \caption{Joint 2D posterior for \chieff vs. detector frame \mc for all the events analyzed in this paper, for Config. A and C. 
    We do not show Config. B to avoid overcrowding the plot. 
    For Config. A (C) the maximum likelihood estimate is indicated with a dot (cross) and the edge of the $90\%$ contour by a solid (dotted) line. }
  \label{Fig.chieff_mc}
\end{figure}

We first present an overview of the results in Fig.~\ref{Fig.chieff_mc}, which shows contours in the \mc--\chieff plane that enclose 90\% of the probability for the seven events discussed in this paper. 
We show Config. A and C, and omit Config. B to avoid overcrowding. Solid (dotted) lines and a dot (cross) mark the contours and the maximum likelihood point for Config. A (C). 
The posteriors on the parameters are formally consistent with each other within their credible intervals, but there are points of difference between the different configurations. 
The differences are relatively minor for most of the events, but notable in the case of GW151216 and GW170403. 
We will therefore first discuss these two cases, and then, we briefly review the properties of the other events, for which results are consistent across the analyses. To better quantify the discrepancies between the different configurations, especially for these two events with large spins, we additionally report the posterior percentile of \chieff values on both tails of the distribution.

\subsection{GW151216}
\label{Sec:GW151216}

\begin{figure}[t]
  \centering
    \includegraphics[width=0.45\textwidth]{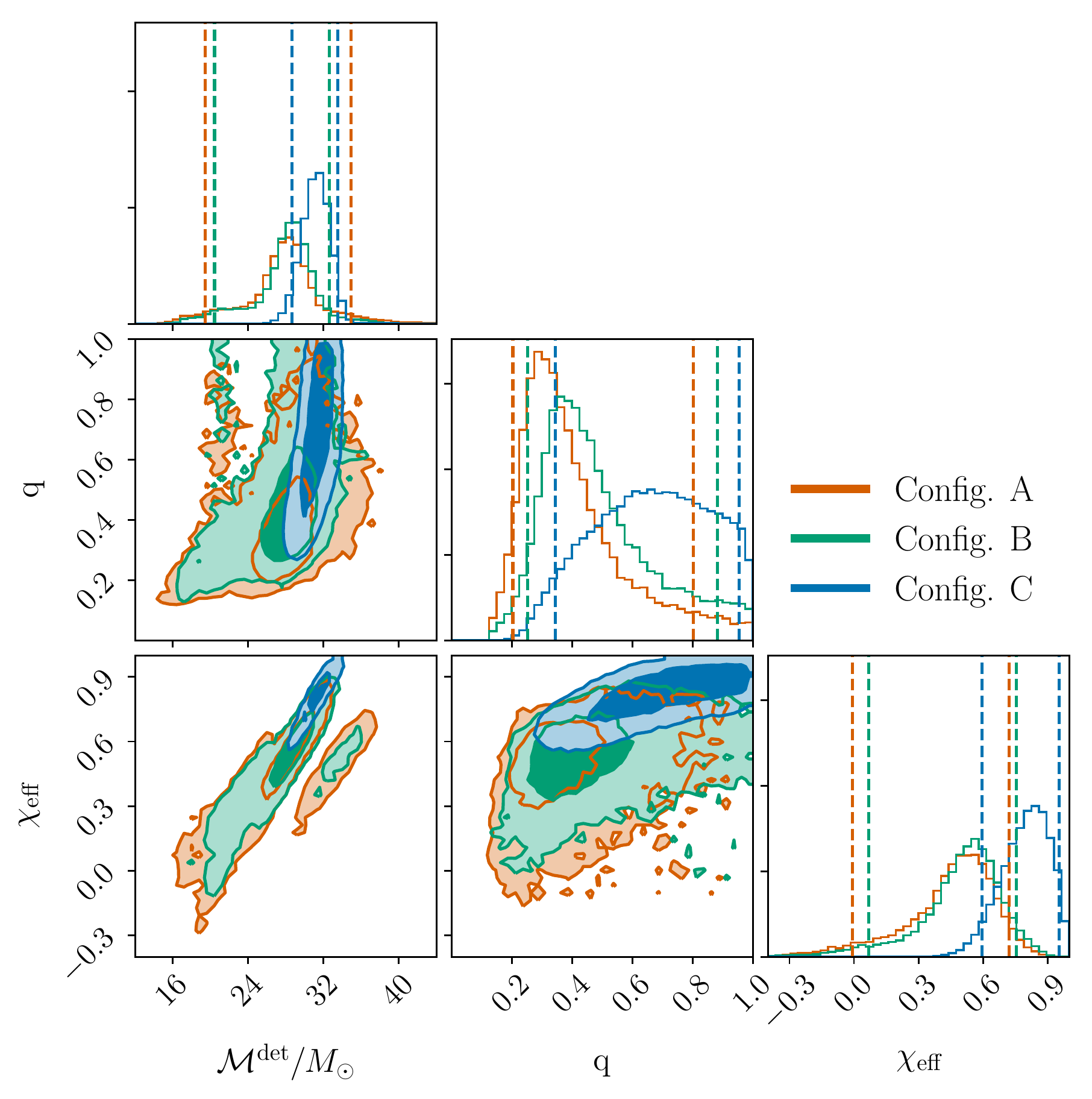}
    \caption{Corner plot for posterior  distributions for GW151216, red for Config. A results, green for Config. B and blue for Config. C.  
    The dashed lines mark the 90\% credible interval, and the dark (light) shaded area marks the 50\%(90\%) contour, same for all corner plots to follow. }
  \label{Fig.corner151216}
\end{figure}

\begin{table}[tb]
\centering
\begin{tabularx}{0.45\textwidth}{@{} bbbb @{}}
\toprule\toprule
 Configuration & A & B &  C \\ [0.5ex] 
\midrule
\mc/\msun & $    28^{+    7}_{-    8} $  & $    29^{+    3}_{-   12} $  & $    31^{+    2}_{-    3} $  \\ 
q & $   0.4^{+  0.5}_{-  0.2} $  & $   0.5^{+  0.4}_{-  0.3} $  & $   0.7^{+  0.3}_{-  0.3} $  \\ 
\chieff & $   0.5^{+  0.2}_{-  0.5} $  & $   0.7^{+  0.2}_{-  0.9} $  & $   0.8^{+  0.1}_{-  0.2} $  \\ 
$D_L$/Gpc & $   1.6^{+  1.3}_{-  0.8} $  & $   1.5^{+  0.4}_{-  0.8} $  & $   2.5^{+  1.2}_{-  1.1} $  \\ 
\midrule
SNR &  8.6&  8.4 &  8.5 \\
$\mathrm{ln}\mathcal{B}_{S/N}$ & 10.8 & 10.6 & - \\
p($\chieff\leq0|{\bf d})$ & 5.2\% & 3.4\% & 0.0\% \\
p($\chieff\geq0.8|{\bf d})$ & 1.3\% & 2.6\% & 52.5\% \\
 \bottomrule\bottomrule
\end{tabularx}
\caption{Properties for GW151216 estimated using 3 different configurations. 
In the upper half, the median values are reported for the source parameters, with error bars marking the span of the 90\% credible intervals. 
The SNRs are calculated using Eq.~\eqref{Eq.SNR}, and the values here corresponds to the maximum likelihood. $\mathrm{ln}\mathcal{B}_{S/N}$ is the natural log Bayes factor. Tables for other events follow the same reporting set-up. p($\chieff\leq0|{\bf d})$ and p($\chieff\geq0.8|{\bf d})$ marks the probability for \chieff to take values less than or equal to 0 and greater or equal to 0.8, respectively.}
\label{Table.Table151216}
\end{table} 

GW151216 was reported by \citet{Zackay:2019tzo} as having a high and positive effective spin, $\chieff=0.8^{+0.1}_{-0.2}$.
Our results are shown in Fig.~\ref{Fig.corner151216} and summarized in Table~\ref{Table.Table151216}.

Of all the events analyzed in this paper, the differences in the inferred parameters across the configurations are the clearest for GW151216.
The chirp mass posteriors obtained from Configs. A and B show fat tails, which are often associated with faint signals such as the ones analyzed here~\cite{Huang:2018tqd}. 
Conversely in Config. C, the distribution of the chirp mass is narrower and centered at $\sim 31~\msun$.
All of the estimates are compatible within their credible intervals.
Similarly, the mass ratio measurements, while having large posterior overlaps, peak at rather different values. 
Configs. A and B have median values of $q = 0.4$ and $q = 0.5$ respectively, whereas Config. C has a median value of $q = 0.7$.
Configs. A and B give marginal, but still non-zero, support for equal mass binaries $(q=1)$.
The median for the luminosity distance is above $1.5$~Gpc for all configurations, with Config. C placing the source at the largest distance among three analyses, $D_L = 2.4^{+1.2}_{-1.1}$~Gpc.

Finally and more importantly, we observe differences in the estimation of \chieff. 
Config. C finds that \chieff is large and positive, while Configs. A and B have low levels of support at zero \chieff.
More specifically, Config. A finds $\chieff = 0.5^{+0.2}_{-0.5}$ and Config. B finds $\chieff = 0.7^{+0.2}_{-0.9}$. 
Both of these posteriors peak at positive values for \chieff, but have long tails extending towards small values. 
The fact that different analyses yield \chieff posteriors that peak at different values can be at least partially explained by the very different priors that are used, see Fig.~\ref{Fig.spin prior}. 
Configs. A and B penalize {\it a priori} large values of \chieff, and thus reduce the prior support at large values.

The $\mathrm{ln}\mathcal{B}_{S/N}$ values of models A and B are similar, with a natural log Bayes factor of 0.2 in favor of Config. A. 
This suggests, as found in~\cite{Zackay:2019tzo}, that there is not enough information available to either support or rule out the presence of spin-precession in GW151216. As an additional test, we repeat the analysis of Config. A while fixing the spins to be zero. 
This provides mild evidence for non-zero spins when compared to our spinning cases, with a natural log Bayes factor of 2.4 (2.2) in favor of precessing (aligned) spins over zero spins. 

It is worth pointing out that any differences in the inferred values of $\mc, q, \chieff, D_L$ across the various analyses are expected to be correlated, as there are significant degeneracies between these parameters \cite{Baird:2012cu, Ng:2018neg, Roulet:2018jbe}.
In the region of parameter space relevant for GW151216, the tightest correlation involves $\mc$, \chieff, $q$. Hence changing the prior on \chieff can affect the inference of the other parameters as well.

We have verified that the choice of waveform models does not play a significant role in the differences by performing a supplementary analysis where all the other analysis choices including the priors and PSD used are the same as Config. B, and only the waveform is varied from SEOBNRv4\_ROM to IMRPhenomD. We find no appreciable difference in the posteriors.

As mentioned in Section \ref{Table.Configuration}, the other points of difference between the configurations are (a) the length of data used, (b) the choice of sampler, (c) the method used to infer the PSD, and consequently, compute the likelihood, and (d) the choice of prior. We performed a number of tests to narrow down the reasons for the discrepancy in the inferred parameters; we present associated details in Appendix~\ref{Appendix}.

In line with our intuition, we find that the most important cause of the differences is the choice of prior: using the `3D isotropic' spin prior causes the mode of the posterior for \chieff to shift to lower values (this is consistent with the analysis in \citet{Zackay:2019tzo}, in which the inference performed using the same prior as in Config. B gives $\chieff = 0.6^{+0.2}_{-0.2}$). None of the other factors (method of PSD estimation, sampler, waveform, segment length, etc.) have as significant an impact on the results.

Apart from the shift in the posteriors for \chieff, there is an additional effect: the tails of the posteriors are systematically broader in Configs. A and B, respectively with 5.2\% and 3.4\% of the \chieff posterior distributions extending below 0 (a similar effect was also reported in Ref.~\cite{Nitz:2019hdf}, with even more dramatic tails in the posteriors,) as compared to 0.0\% of Config. C.
Configs. A and B also show significantly less support at high \chieff values, 1.3\% and 2.6\% above \chieff = 0.8 compared to 52.5\% for Config. C. Deeper investigation shows that this additional phenomenon is related to a combination of the sampler used, and the treatment of spectral lines in the data while calculating the likelihood. Further details can be found in Appendix~\ref{Appendix}. In light of these investigations, we conclude that whether the \chieff posterior of GW151216 is an outlier compared to the other systems we discuss in this paper, Fig.~\ref{Fig.chieff_mc}, depends strongly on the details of the analysis.

Config. A recovers SNR of 8.6, and Config. B (C) recover a similar SNR of $\sim 8.4(8.5)$. The $\mathrm{ln}\mathcal{B}_{S/N}$ values are also comparable, 10.8 and 10.6 from Configs. A and B.

\subsection{GW170403}

\begin{figure}[t]
  \centering
    \includegraphics[width=0.45\textwidth]{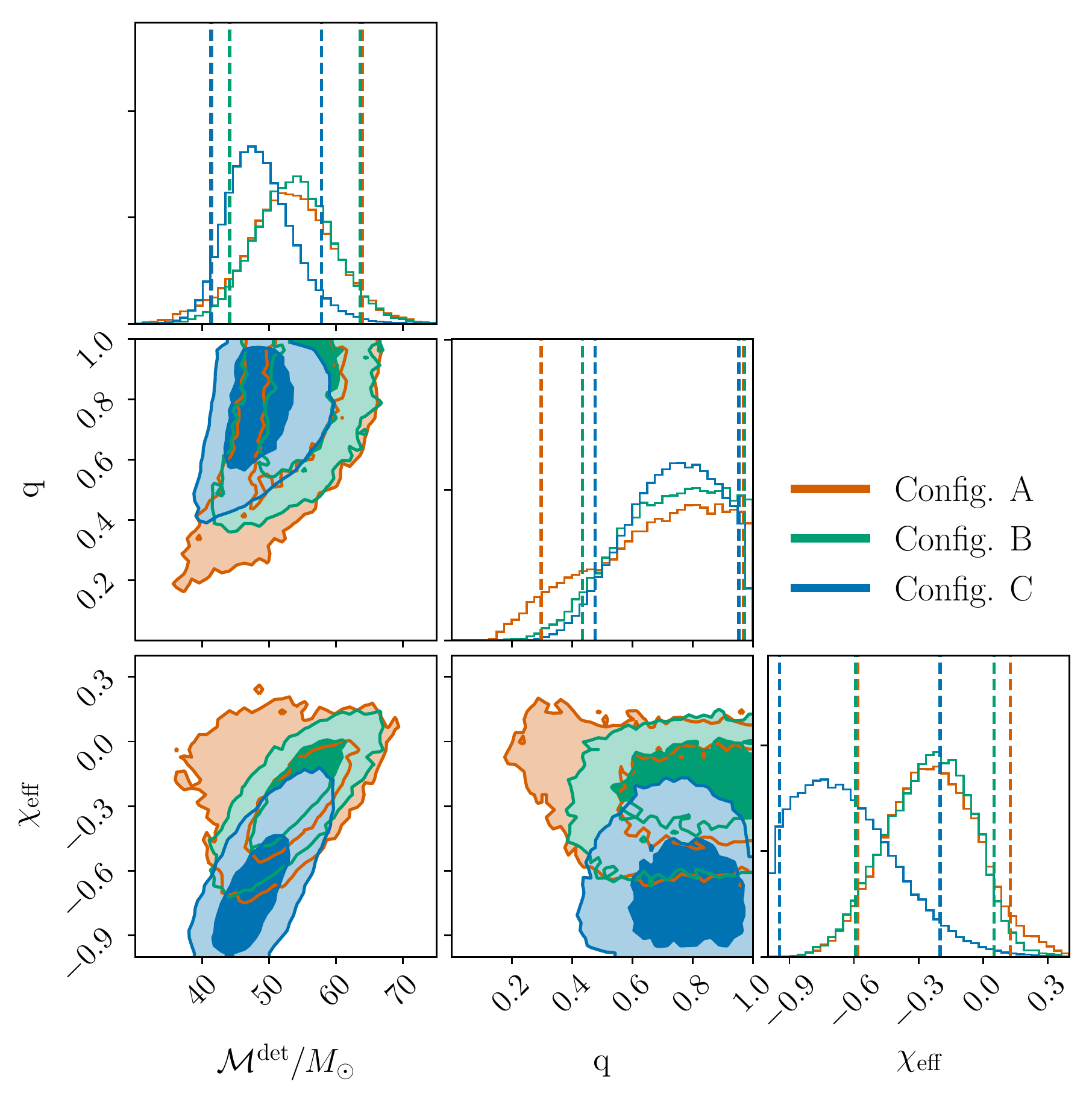}
    \caption{Corner plot for posterior distributions for GW170403, red for Config. A results, green for Config. B and blue for Config. C.  }
  \label{Fig.corner170403}
\end{figure}

\begin{table}[tb]
\centering
\begin{tabularx}{0.45\textwidth}{@{} bbbb @{}}
\toprule\toprule
 Configuration & A & B & C \\ [0.5ex] 
\midrule
\mc/\msun & $    53^{+   11}_{-   12} $  & $    54^{+   10}_{-   10} $  & $    48^{+   10}_{-    7} $  \\ 
q & $   0.7^{+  0.3}_{-  0.4} $  & $    0.7^{+  0.2}_{-  0.3} $  & $   0.7^{+  0.2}_{-  0.3} $  \\

\chieff & $  -0.2^{+  0.4}_{-  0.3} $  & $  -0.2^{+  0.3}_{-  0.4} $  & $  -0.7^{+  0.5}_{-  0.3} $  \\ 

$D_L$/Gpc & $   2.8^{+  2.3}_{-  1.5} $  & $   3.2^{+  2.2}_{-  1.6} $  & $   2.7^{+  1.5}_{-  1.2} $  \\ 
\midrule
SNR &  8.4&  8.1 &  8.2 \\
$\mathrm{ln}\mathcal{B}_{S/N}$ & 11.8 & 11.2 & - \\
p(${\chieff\leq-0.7|{\bf d})}$ & 1.2\% & 1.4\% & 42.7\% \\
p($\chieff\geq0|{\bf d})$ & 14.0\% & 9.6\% & 1.2\% \\
\bottomrule\bottomrule
\end{tabularx}
\caption{Properties for GW170403 estimated using 3 different configurations. p(${\chieff\leq-0.7|{\bf d})}$ and p($\chieff\geq0|{\bf d})$ marks the probability for \chieff to take values less than or equal to -0.7 and greater or equal to 0, respectively. }
\label{Table.Table170403}
\end{table} 

In \citet{Venumadhav:2019lyq}, GW170403 was found to have large and negative \chieff.
With Config. C we find $\chieff=-0.7^{+ 0.5}_{- 0.3}$, which excludes $\chieff=0$ from the 90\% credible interval, with 1.2\% of the \chieff posterior distributions extending above 0, and 42.7\% below -0.7. Here we also find that Configs. A and B yield posteriors with larger support at $\chieff=0$, Tab.~\ref{Table.Table170403}. Config. A ($\chieff= -0.2^{+ 0.4}_{- 0.3}$) and Config. B ($-0.2^{+ 0.3}_{- 0.4} $) have 14.0\%(1.2\%) and 9.6\%(1.4\%) of the \chieff posterior distributions extending above 0 (below -0.7), respectively.
 
Fig.~\ref{Fig.corner170403} shows a difference between aligned and precessing waveform models, most clearly seen in the joint $q$--$\chieff$ posterior. 
The precessing degrees of freedom allowed in Config. A to alter the well-known correlation between \chieff and mass ratio~\cite{Baird:2012cu,Ng:2018neg}, yielding a broader posterior distribution for the mass ratio, whose lower end of the 90\% credible interval now reaches $\sim 0.3$. 
Again, we do not find enough information present to confirm or rule out the presence effects of spin-induced orbital precession.

However, the posteriors for $q$, chirp mass and luminosity distance are more consistent across our analysis configurations for GW170403 than they are for GW151216.
This suggest that the small differences we observe for GW170403 can be entirely or nearly entirely explained by the different priors used for \chieff, which ``push'' the posteriors in Configs. A and B closer to 0.
Similar to the findings in~\cite{Vitale:2017cfs}, varying the prior choices has a more substantial effect on low-SNR observations like the events analyzed in this paper, so general caution should be exercised when drawing astrophysical inferences using quantities that are as strongly prior dependent as the \chieff measurements presented here.
On the other hand, the comparison for GW151216 suggests that for low-SNR events the specific realization of the noise and the detector behavior around the trigger time may amplify the differences introduced by the PSD estimation, which is quite different in Configs. A and B compared to Config. C.

We notice that Config. C has a more pronounced tail at negative \chieff, resulting in stronger support at lower values of the chirp mass.
Systems with more negative \chieff produce shorter GW signals~\cite{Ng:2018neg}. 
This can be roughly compensated for by decreasing the chirp mass~\cite{Roulet:2018jbe}.

\subsection{GW170202} 

\begin{figure}[t]
  \centering
    \includegraphics[width=0.45\textwidth]{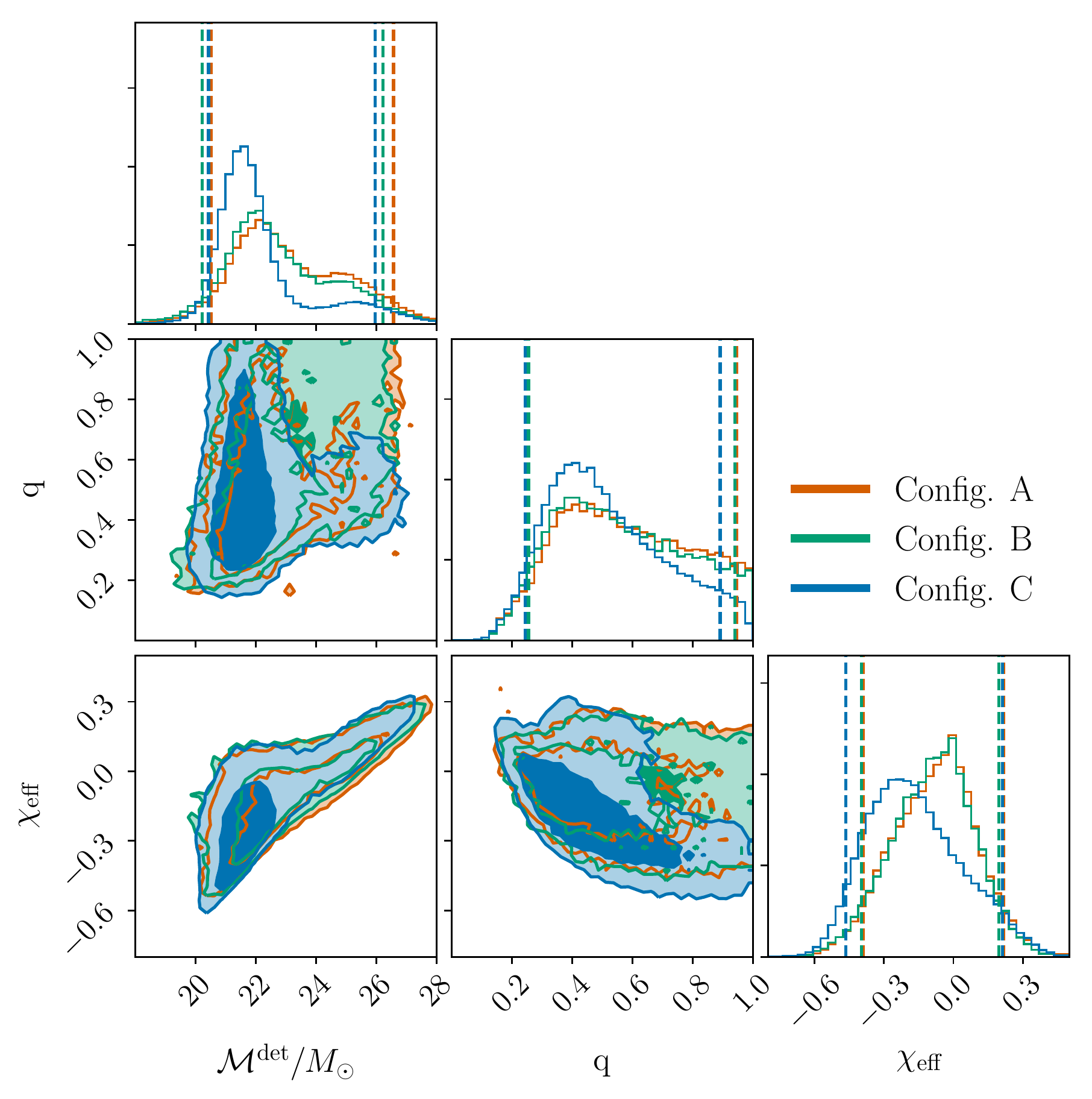}
    \caption{Corner plot for posterior distributions for GW170202, red for Config. A results, green for Config. B and blue for Config. C.}
  \label{Fig.corner170202}
\end{figure}

\begin{table}[tb]
\centering
\begin{tabularx}{0.45\textwidth}{@{} bbbb @{}}
\toprule\toprule
 Configuration & A & B &  C \\ [0.5ex] 
\midrule
\mc/\msun & $    23^{+    4}_{-    2} $  & $    23^{+    4}_{-    2} $  & $    22^{+    4}_{-    1} $  \\ 
q & $   0.6^{+  0.4}_{-  0.3} $  & $    0.5^{+  0.4}_{-  0.3} $  & $   0.5^{+  0.4}_{-  0.2} $  \\ 
\chieff & $  -0.1^{+  0.3}_{-  0.3} $  & $  -0.1^{+  0.3}_{-  0.3}  $  & $  -0.2^{+  0.4}_{-  0.3} $  \\ 
$D_L$/Gpc & $   1.5^{+  1.1}_{-  0.8} $  & $    1.5^{+  1.0}_{-  0.8} $  & $   1.5^{+  0.8}_{-  0.6} $  \\ 
\midrule
SNR &  8.5&  8.3 &  8.5 \\
$\mathrm{ln}\mathcal{B}_{S/N}$ & 10.5 & 10.9 & - \\
\bottomrule\bottomrule
\end{tabularx}
\caption{Properties for GW170202 estimated using 3 different configurations.  }
\label{Table.Table170202}
\end{table} 

All the analyses yield consistent results for GW170202, as seen in Tab.~\ref{Table.Table170202} and Fig.~\ref{Fig.corner170202}.
The detector frame chirp mass for GW170202 is estimated to be $22^{+4}_{-1}~\msun$ ($23^{+4}_{-2} ~\msun$) with flat-in-\chieff prior (or otherwise).  
Of all the sources we discuss in this work, GW170202 is the one for which we measure the lowest mass ratio, consistently across the configurations: Config. A yields $q= 0.6^{+ 0.4}_{- 0.3}$, while Configs. B and C have an even lower median, $q= 0.5^{+ 0.4}_{- 0.3}$ and $q= 0.5^{+ 0.4}_{- 0.2}$, respectively. The \chieff posterior is broadly consistent across the three analyses, with Configs. A and B peaking closer to zero, as expected given that their priors prefer values closer to zero.

\subsection{GW170121, GW170304,  GW170425, GW170727}

\begin{figure}[h]
  \centering
    \includegraphics[width=0.45\textwidth]{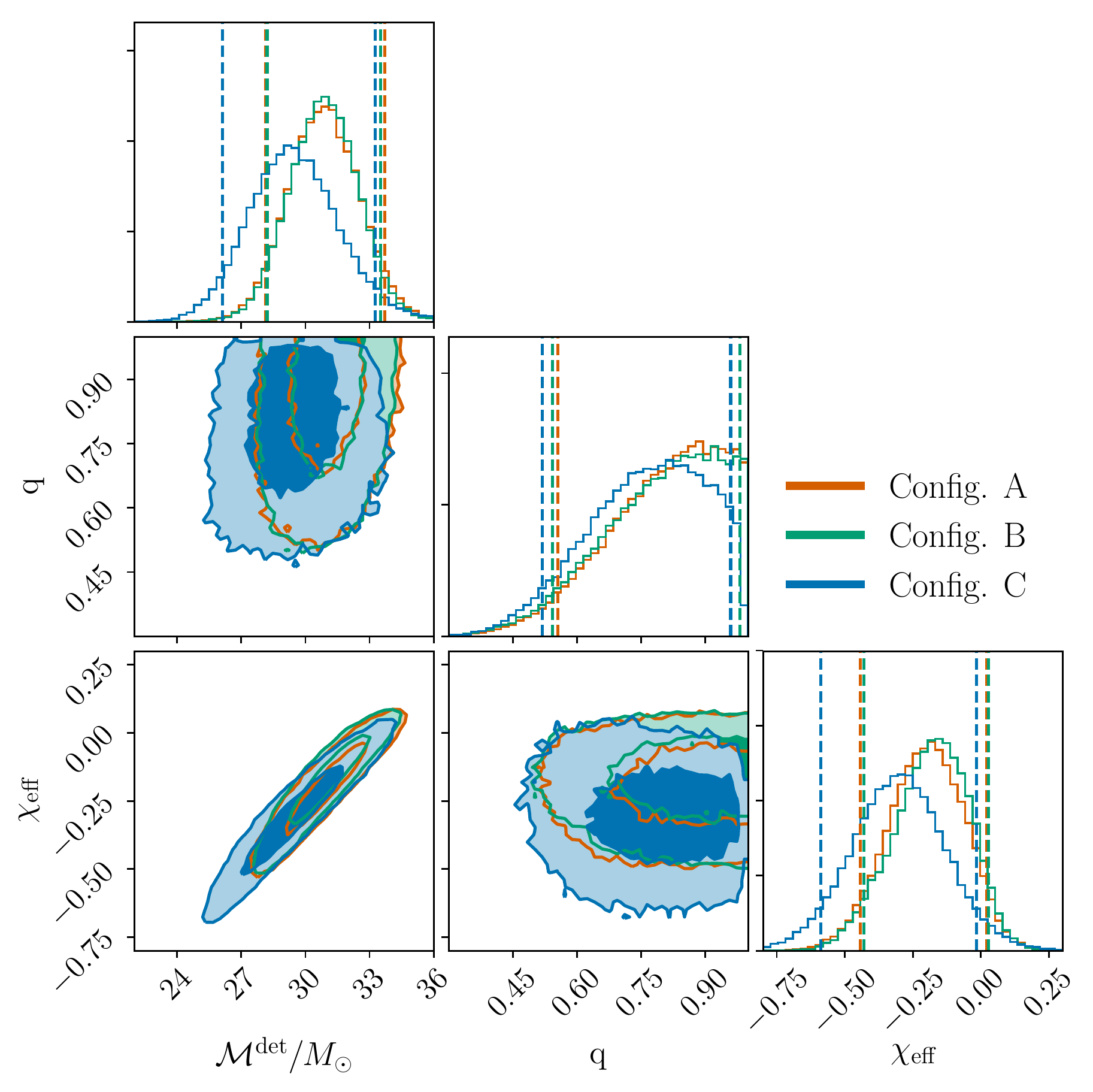}
    \caption{Corner plot for posterior  distributions for GW170121, red for Config. A results, green for Config. B and blue for Config. C.}
  \label{Fig.corner170121}
\end{figure}

\begin{table}[h!]
\centering
\begin{tabularx}{0.45\textwidth}{@{} bbbb @{}}
\toprule\toprule
 Set-up & A & B &  C \\ [0.5ex] 
\midrule
\mc/\msun & $    31^{+    3}_{-    3} $  & $    31^{+    3}_{-    3} $  & $    29^{+    4}_{-    3} $  \\ 
q & $   0.8^{+  0.2}_{-  0.3} $  & $   0.8^{+  0.2}_{-  0.3} $  & $   0.8^{+  0.2}_{-  0.3} $  \\ 
\chieff & $  -0.2^{+  0.2}_{-  0.2} $  & $  -0.2^{+  0.2}_{-  0.3} $  & $  -0.3^{+  0.3}_{-  0.3} $  \\ 
$D_L$/Gpc & $   1.3^{+  0.9}_{-  0.7} $  & $   1.3^{+  0.9}_{-  0.8} $  & $   1.3^{+  0.9}_{-  0.7} $  \\ 
\midrule
SNR &  10.8&  10.7 &  10.9 \\
$\mathrm{ln}\mathcal{B}_{S/N}$ & 30.9 & 31.1 & - \\
\bottomrule\bottomrule
\end{tabularx}
\caption{Properties for GW170121 estimated using 3 different configurations. }
\label{Table.Table170121}
\end{table}

\begin{figure}[t]
  \centering
    \includegraphics[width=0.45\textwidth]{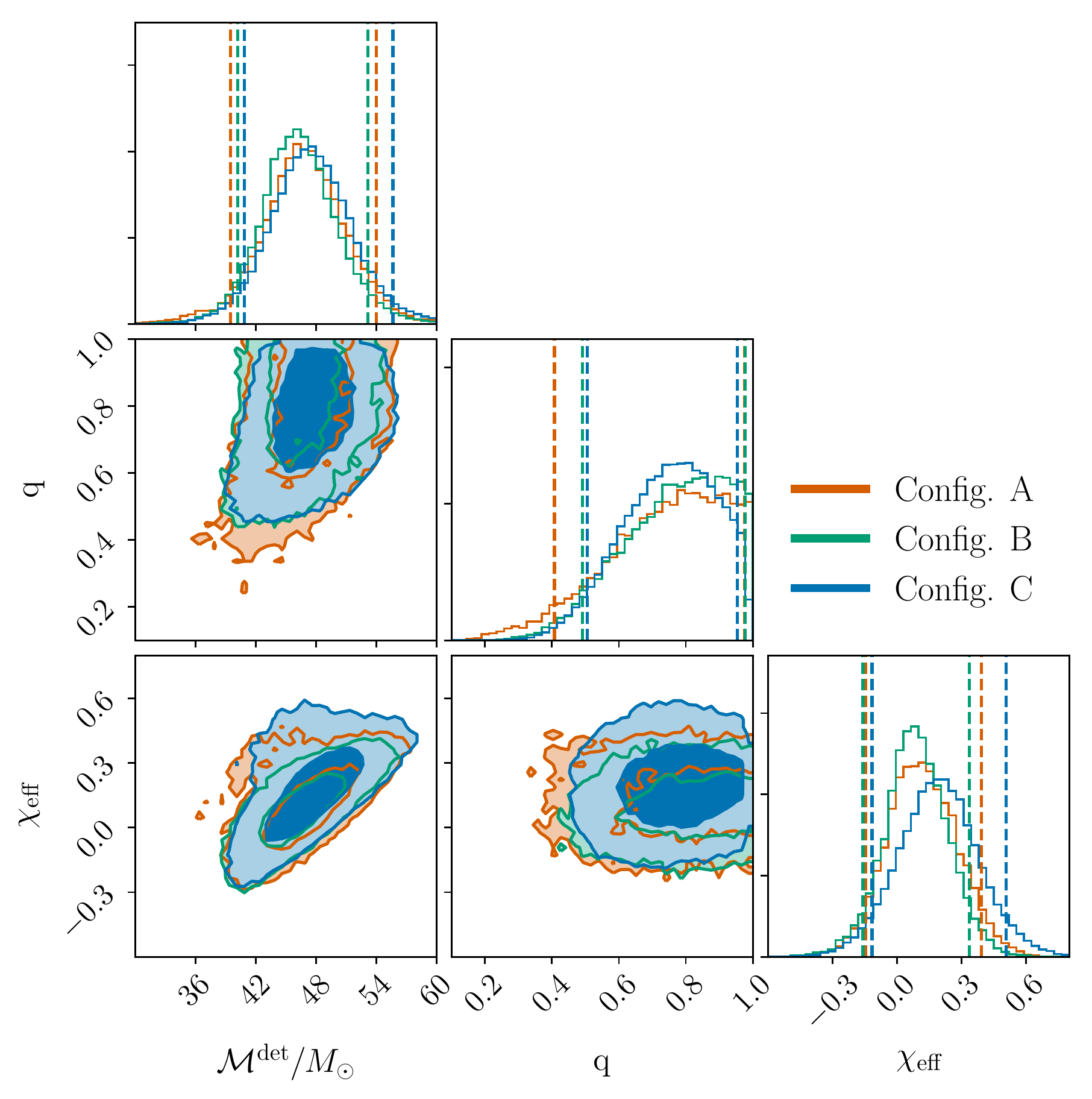}
    \caption{Corner plot for posterior distributions for GW170304, red for Config. A results, green for Config. B and blue for Config. C.  }
  \label{Fig.corner170304}
\end{figure}

\begin{table}[tb]
\centering
\begin{tabularx}{0.45\textwidth}{@{} bbbb @{}}
\toprule\toprule
 Configuration & A &B &  C \\ [0.5ex] 
\midrule
\mc/\msun & $    47^{+    7}_{-    7} $  & $    46^{+    7}_{-    6} $  & $    48^{+    8}_{-    7} $  \\ 
q & $   0.8^{+  0.2}_{-  0.4} $  & $   0.8^{+  0.2}_{-  0.3} $  & $   0.8^{+  0.2}_{-  0.3} $  \\ 
\chieff & $   0.1^{+  0.3}_{-  0.3} $  & $   0.1^{+  0.2}_{-  0.2} $  & $   0.2^{+  0.3}_{-  0.3} $  \\ 
$D_L$/Gpc & $   2.7^{+  1.6}_{-  1.4} $  &  $   2.6^{+  1.6}_{-  1.4} $  & $   3.0^{+  1.6}_{-  1.3} $  \\ 
\midrule
SNR &  9.0&  8.7 &  8.7 \\
$\mathrm{ln}\mathcal{B}_{S/N}$ & 16.0 & 15.7 & - \\
\bottomrule\bottomrule
\end{tabularx}
\caption{Properties for GW170304 estimated using 3 different configurations. }
\label{Table.Table170304}
\end{table}

GW170121 has the highest SNR among the events discussed in the paper. 
The PE results are consistent with a heavy, near equal-mass BBH with a preference for negative values of \chieff, at luminosity distance of $\sim 1.3$~Gpc.
Our results are shown in Fig.~\ref{Fig.corner170121} and summarized in Table~\ref{Table.Table170121}. 

\begin{figure}[tb]
  \centering
    \includegraphics[width=0.45\textwidth]{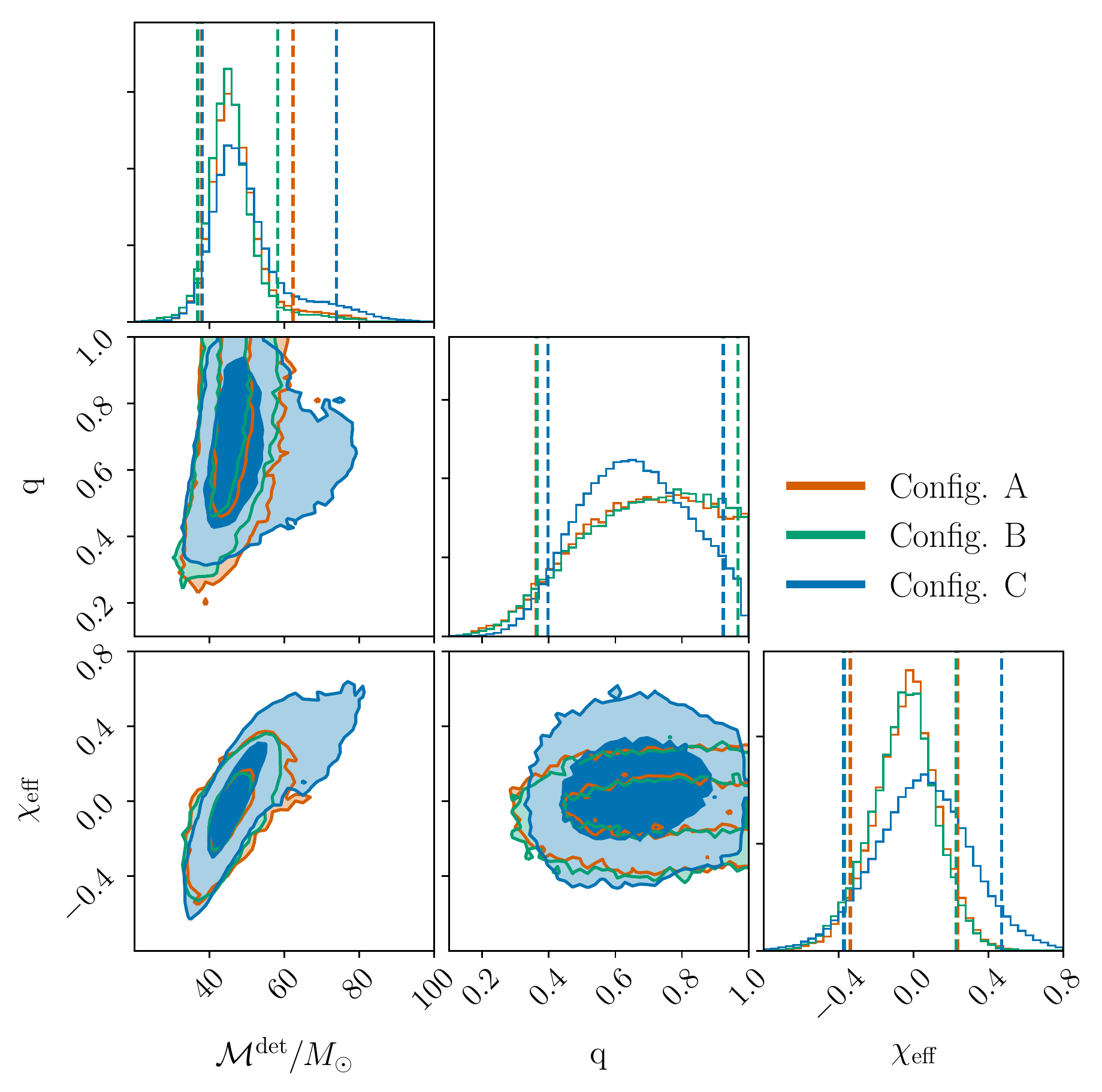}
    \caption{Corner plot for posterior distributions for GW170425, red for Config. A results, green for Config. B and blue for Config. C.}
  \label{Fig.corner170425}
\end{figure}

GW170304 and GW170425 are similar systems, with detector frame chirp masses of $\sim 47~\msun$, \chieff posteriors centered near zero, luminosity distance of $\sim 3$~Gpc and a preference for nearly equal masses, as shown in Figs.~\ref{Fig.corner170304} and~\ref{Fig.corner170425}, and Tables~\ref{Table.Table170304} and~\ref{Table.Table170425}.

While small differences are seen across the configurations, the posteriors obtained from the three analyses are all broadly consistent, and depict very similar results: GW170304 and GW170425 are broadly similar to the majority of the BBHs discovered in LIGO--Virgo data: massive systems with nearly equal component masses and (apparent) \chieff values consistent with zero. 
These heavy BBHs may arise from a common formation scenario~\cite{LIGOScientific:2018mvr,LIGOScientific:2018jsj}.%

\begin{table}[tb]
\centering
\begin{tabularx}{0.45\textwidth}{@{} bbbb @{}}
\toprule\toprule
 Configuration & A & B &  C \\ [0.5ex] 
\midrule
\mc/\msun & $    46^{+   16}_{-    8} $  & $    45^{+   13}_{-    8} $  & $    48^{+   26}_{-   10} $  \\ 

q & $   0.7^{+  0.3}_{-  0.3} $  & $   0.7^{+  0.3}_{-  0.3} $  & $   0.7^{+  0.3}_{-  0.3} $  \\ 

\chieff & $  0.0^{+  0.3}_{-  0.3} $  & $  0.0^{+  0.3}_{-  0.3} $  & $   0.1^{+  0.4}_{-  0.4} $  \\ 

$D_L$/Gpc & $   2.8^{+  2.0}_{-  1.4} $  & $   2.7^{+  1.9}_{-  1.4} $  & $   3.3^{+  2.9}_{-  1.6} $  \\ 
\midrule
SNR &  8.4&  8.4 &  8.0\\
$\mathrm{ln}\mathcal{B}_{S/N}$ & 14.2 & 14.3 & - \\

\bottomrule\bottomrule
\end{tabularx}
\caption{Properties for GW170425 estimated using 3 different configurations. }
\label{Table.Table170425}
\end{table}

\begin{figure}[tb]
  \centering
    \includegraphics[width=0.45\textwidth]{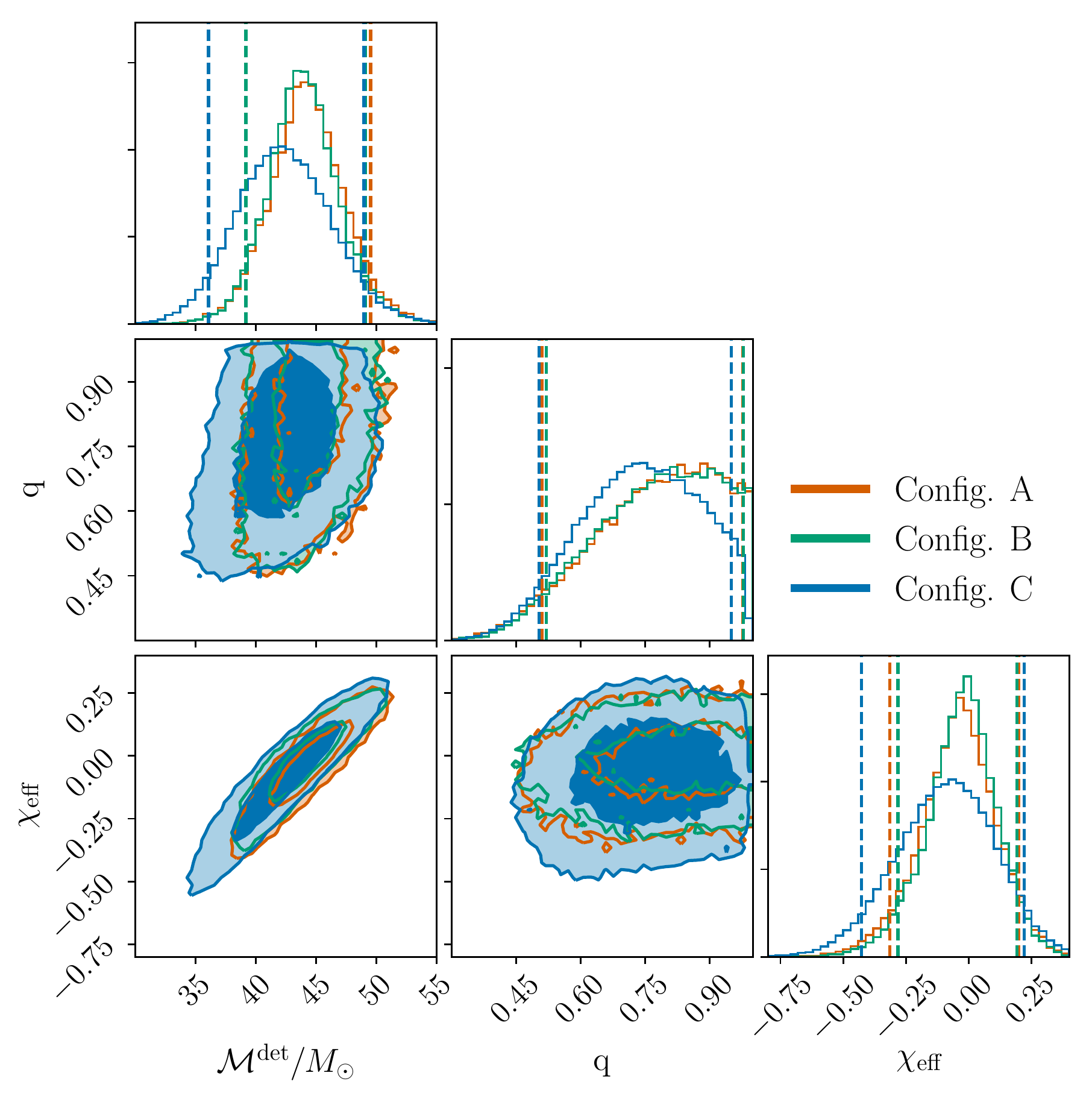}
    \caption{Corner plot for posterior distributions for GW170727, red for Config. A results, green for Config. B and blue for Config. C.  }
  \label{Fig.corner170727}
\end{figure}

The same is true for GW170727, as seen in Tab.~\ref{Table.Table170727} and Fig.~\ref{Fig.corner170727}, which appears only slightly less massive %
and closer, at a recovered median distance of $\sim 2.5$~Gpc.
For all configurations, the \chieff posterior is centered around zero. 
It is worth stressing that even though the SNR reported for Config. A in  Table~\ref{Table.Table170727} is 10\% higher than that for Config. B, we do not find significant evidence in support of the precessing model, with the two configurations yielding similar Bayes factors. The reason is that the SNRs we report are calculated at the point in parameter space that yields the maximum likelihood, whereas the Bayesian evidence, and hence the Bayes factors, are integrated over the whole parameter space, Eq.~\ref{Eq.Evidence}. The median SNR is thus a better tracer for the evidence. We indeed find that the median SNRs of Configs. A and B only differ by a fraction of a percent.

\begin{table}[tb]
\centering
\begin{tabularx}{0.45\textwidth}{@{} bbbb @{}}
\toprule\toprule
 Configuration & A  & B &  C \\ [0.5ex] 
\midrule
\mc/\msun & $    44^{+    5}_{-    5} $  & $    44^{+    5}_{-    5} $  & $    42^{+    7}_{-    6} $  \\ 

q & $   0.8^{+  0.2}_{-  0.3} $  & $   0.8^{+  0.2}_{-  0.3} $  & $   0.7^{+  0.2}_{-  0.2} $  \\ 

\chieff & $  -0.0^{+  0.2}_{-  0.3} $  & $  -0.0^{+  0.2}_{-  0.3} $  & $  -0.1^{+  0.3}_{-  0.3} $  \\ 

$D_L$/Gpc & $   2.4^{+  1.3}_{-  1.2} $  & $   2.4^{+  1.3}_{-  1.2} $  & $   2.5^{+  1.3}_{-  1.1} $  \\ 
\midrule
SNR &  10.2&  9.4 &  9.0\\

$\mathrm{ln}\mathcal{B}_{S/N}$ & 22.6 & 22.5 & - \\
\bottomrule\bottomrule
\end{tabularx}
\caption{Properties for GW170727 estimated using 3 different configurations.  }
\label{Table.Table170727}
\end{table}

\section{Conclusions}\label{Sec.Conclusions}

We report on a comparison of the source property measurement  of the 7 BBH sources first presented in~\cite{Zackay:2019tzo,Venumadhav:2019lyq}.
The analysis therein (corresponding to Config. C in this paper) includes two binaries with \chieff significantly deviating from zero. 
We also perform parameter estimation analyses using the standard algorithms of the LVC \cite{LIGOScientific:2018mvr, TheLIGOScientific:2016wfe, TheLIGOScientific:2016pea} where we use both waveform models allowing for spin-precession (Config. A) and assuming spins only (anti-)aligned to the orbital angular momentum (Config. B).
In the analysis for Configs. A and B, the data from the GW detectors is assumed to be described by a stationary and Gaussian noise process modelled on 4-second-long data segment under analysis using the spectral model in \textsc{BayesWave}, whereas Config. C assumes the noise to be well described by a PSD estimated through Welch's method from a significantly longer data segment surrounding the GW signal times a time-dependent normalization that is measured on a $\sim15$\,second scale.
The three configurations also differ significantly in their respective prior assumptions on the black hole spin parameters, which as shown by~\cite{Vitale:2017cfs} could have a significant effect on the inferred posterior distributions, especially for high-mass BBH systems with low SNRs such as the ones presented in this study.

Configs. A and B recover lower values for \chieff, as compared to~\cite{Zackay:2019tzo,Venumadhav:2019lyq}, which becomes consistent with zero or with low component spins for all reported BBHs, Fig.~\ref{Fig.chieff_mc}.
In particular, for GW151216, Ref.~\cite{Zackay:2019tzo} reported a positive \chieff of $0.8^{+0.1}_{-0.2}$, while we report $\chieff = 0.5^{+0.2}_{-0.5}$ and $\chieff = 0.7^{+0.2}_{-0.9}$ (corresponding to Config. C, A and B, respectively), and similarly for GW170403, Ref.~\cite{Zackay:2019tzo} reported a negative \chieff of $-0.7^{+0.5}_{-0.3}$, while we report $\chieff = -0.2^{+0.4}_{-0.3}$ and $\chieff = -0.2^{+0.3}_{-0.4}$ (again corresponding to Config. C, A, B).

Ultimately, one should choose priors that reflect the underlying population of black holes. 
In order to measure the mass and spin distributions of this population, the prior choices applied for any individual event must be removed, so to not double count the prior probability impact, and the ``raw'' likelihood distributions used to infer the properties of the population~\cite{LIGOScientific:2018jsj,LIGOScientific:2018mvr}. 
Such an analysis was recently carried out in~\cite{Galaudage:2019jdx}, whose population-informed posteriors broadly agree with those derived in this work.
Future parameter estimation analysis will benefit from the use of population-informed priors, especially as the number of detected GW events grows.

\section{Acknowledgments}
The authors would like to thank Roberto Cotesta for useful discussions. Y.~H., C.-~J.~H., and S.~V.~acknowledge support of the MIT physics department through the Solomon Buchsbaum Research Fund, the National Science Foundation, and the LIGO Laboratory. 
A.~Z.~is funded by NSF Grant PHY-1912578.
J.~R.~thanks the Center for Computational Astrophysics for hospitality.
T.~V. and L.~D. acknowledge the support of John Bahcall Fellowships at the Institute for Advanced Study.
B.~Z.~is supported by the Frank and Peggy Taplin membership fund.
M.Z. is supported by NSF grants AST1409709, PHY-1820775 the Canadian
Institute for Advanced Research (CIFAR) program on
Gravity and the Extreme Universe and the Simons Foundation Modern Inflationary Cosmology initiative.
The authors  acknowledge usage of LIGO Data Grid clusters.
This research has made use of data, software and/or web tools obtained from the Gravitational Wave Open Science Center (https://www.gw-openscience.org), a service of LIGO Laboratory, the LIGO Scientific Collaboration and the Virgo Collaboration. 
LIGO was constructed by the California Institute of Technology and Massachusetts Institute of Technology with funding from the National Science Foundation and operates under cooperative agreement PHY-0757058.
Virgo is funded by the French Centre National de Recherche Scientifique (CNRS), the Italian Istituto Nazionale della Fisica Nucleare (INFN) and the Dutch Nikhef, with contributions by Polish and Hungarian institutes.
This is LIGO Document Number DCC-P2000082.

\appendix
\section{Further investigation on GW151216}\label{Appendix}

Among the events which are discussed in this paper, GW151216 shows the biggest differences in its inferred parameters between the three setups. We summarized the differences in Section \ref{Sec:GW151216} in the body of the paper. In this appendix, we present the results of a deeper investigation into the possible causes. As we mentioned in Section \ref{Method}, the configurations used in this study differ along five axes: (a) the segment of data used, (b) the algorithm used to estimate the PSD, and consequently, compute the likelihood $\lk(d | \boldsymbol{\theta})$, (c) the sampler used in parameter estimation, (d) the waveform model, and finally (e) the prior on the spins of the black holes. It is not practical to explore every combination of factors, and hence, we perform a few controlled experiments by varying the choices that we expect to be the most important. We checked that the different choices of the waveform model made no difference in this case, so we omit that factor from the rest of the discussion.

From the results in Ref.~\cite{Zackay:2019tzo} as well as Section \ref{Sec:GW151216} of the main body for this paper, we expect that the choice of prior can play a significant role. 
The configurations in Table~\ref{Table.Configuration} differ in both the prior and other analysis choices. Hence it is worthwhile to fix the prior and vary the other choices.
    
\begin{figure*}[!t]
    \centering
    \begin{subfigure}[t]{0.45\textwidth}
        \centering
        \includegraphics[width=\linewidth]{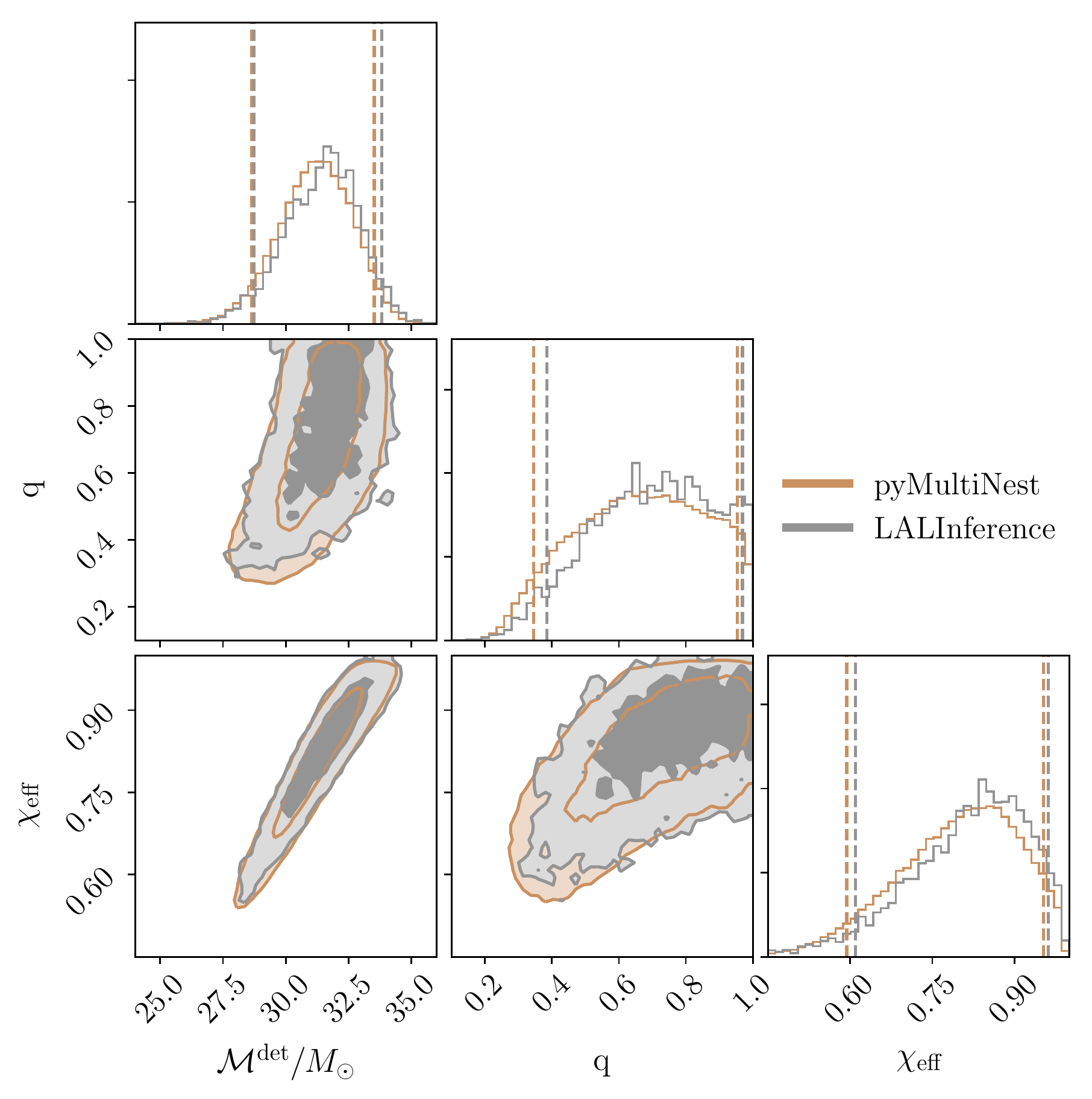} 
        \caption{Effect of choice of sampler} \label{Fig.SamplerTest}
    \end{subfigure}
    \hfill    
    \begin{subfigure}[t]{0.45\textwidth}
        \centering
        \includegraphics[width=\linewidth]{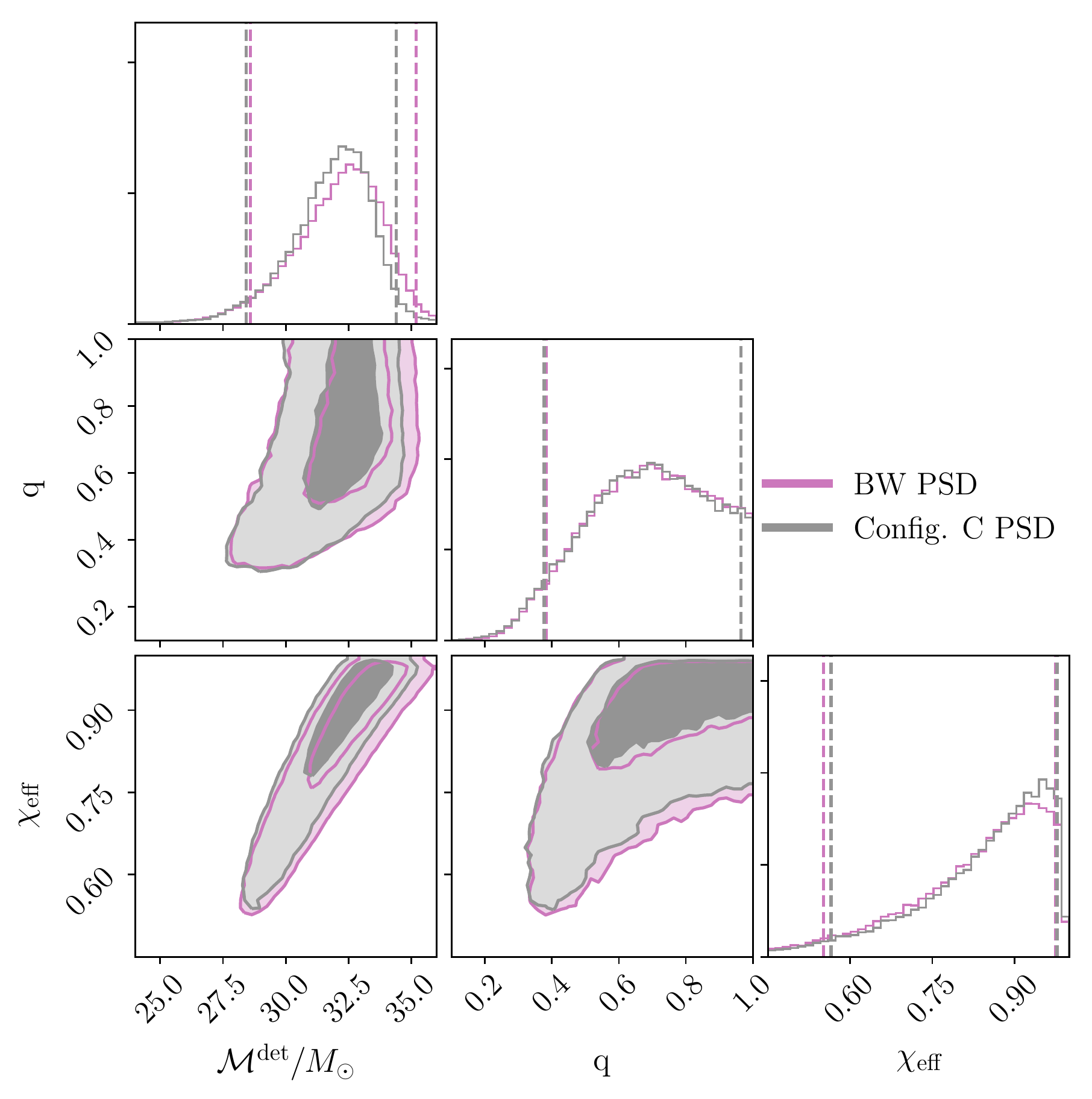} 
        \caption{Effect of algorithm used to estimate PSD} \label{Fig.PSDTest}
    \end{subfigure}
    \caption{\label{Fig.Flatchieffcomparison} Corner plot for posterior distributions for GW151216, using flat-in-\chieff spin prior, IMRPhenomD, and a 128 second long data segment. The left and right panels, respectively, show the effects of successively changing the sampler and the method of PSD estimation from those of Config. C to those of Configs. A and B. The gray filled contours show the posterior using the sampler in \linf, and the PSD estimated using Welch's method (with a drift correction factor, used in Config. C). On the left panel, the brown contours mark the same analysis done with the \textsc{pyMultinest}, and on the right, the purple contours show the effect of changing only the PSD to the one estimated using \textsc{BayesWave} (also used in Configs. A\& B). This figure shows that under the flat-in-\chieff prior, the inference is insensitive to other analysis choices.}
\end{figure*}

We first restrict to the flat-in-\chieff spin prior of Config. C. 
Figure \ref{Fig.Flatchieffcomparison} shows the effect of successively changing the sampler and the method of PSD estimation from those of Config. C to those of Configs. A and B. 
Firstly, Fig.~\ref{Fig.SamplerTest} shows the effect of varying the samplers, \textsc{pyMultiNest} and \linf, while keeping the rest of the configuration identical to Config. C. 
Next, Fig.~\ref{Fig.PSDTest} shows the impact of varying the method used to estimate the PSD from Welch's method to \textsc{BayesWave}. 
We see that the posteriors are identical to those of Config. C, which implies that under the flat-in-\chieff prior, the rest of the analysis choices do not significantly impact parameter inference, and consequently, the only way to go to the results of Config. B (and A) is to choose a different spin prior.

\begin{figure*}[!t]
    \centering
    \begin{subfigure}[t]{0.45\textwidth}
    	\includegraphics[width=\linewidth]{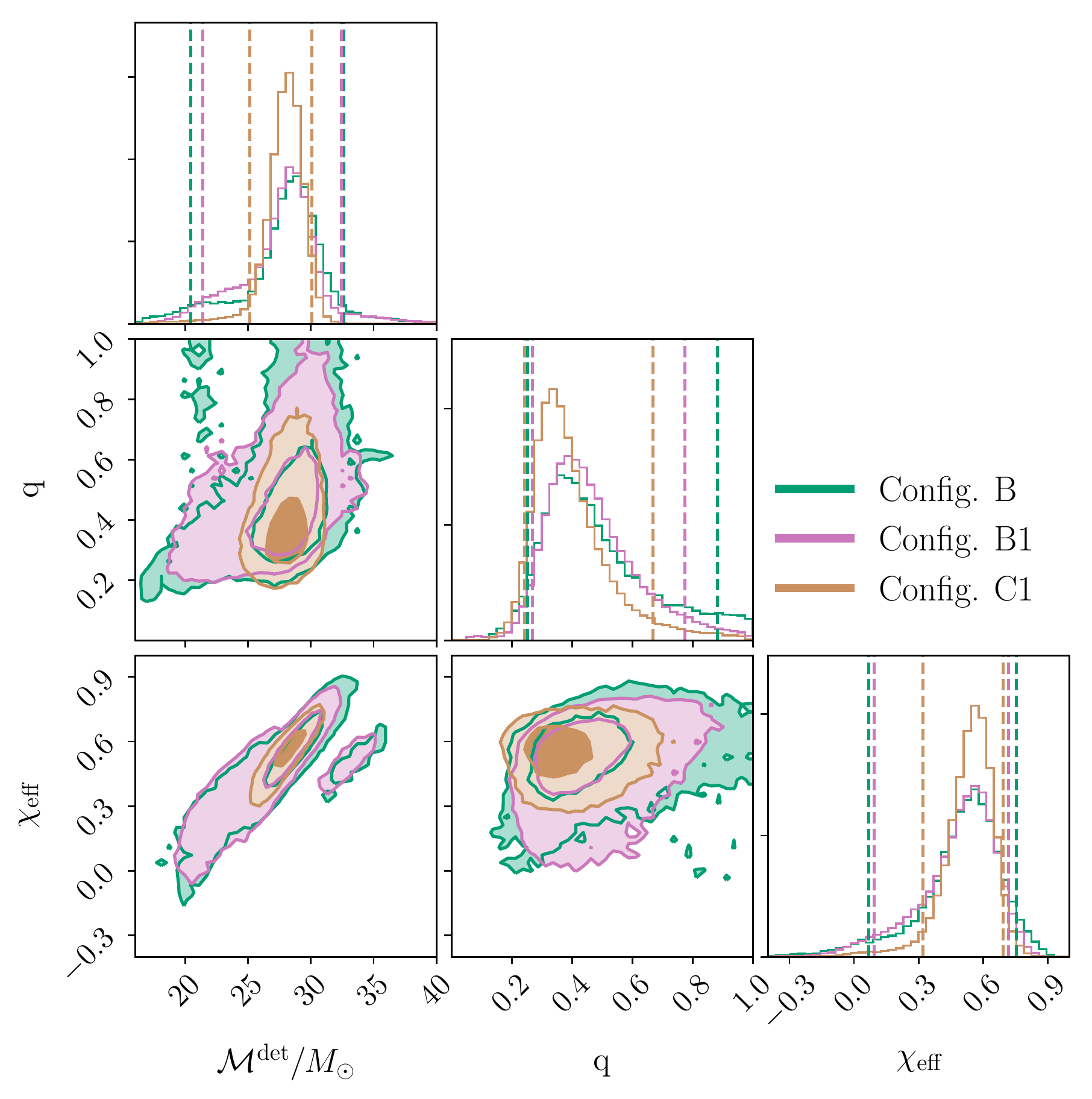}
        \caption{Effect of choice of sampler} \label{Fig.SamplerTest_isotropic}
    \end{subfigure}
    \hfill
    \begin{subfigure}[t]{0.45\textwidth}
    	\includegraphics[width=\linewidth]{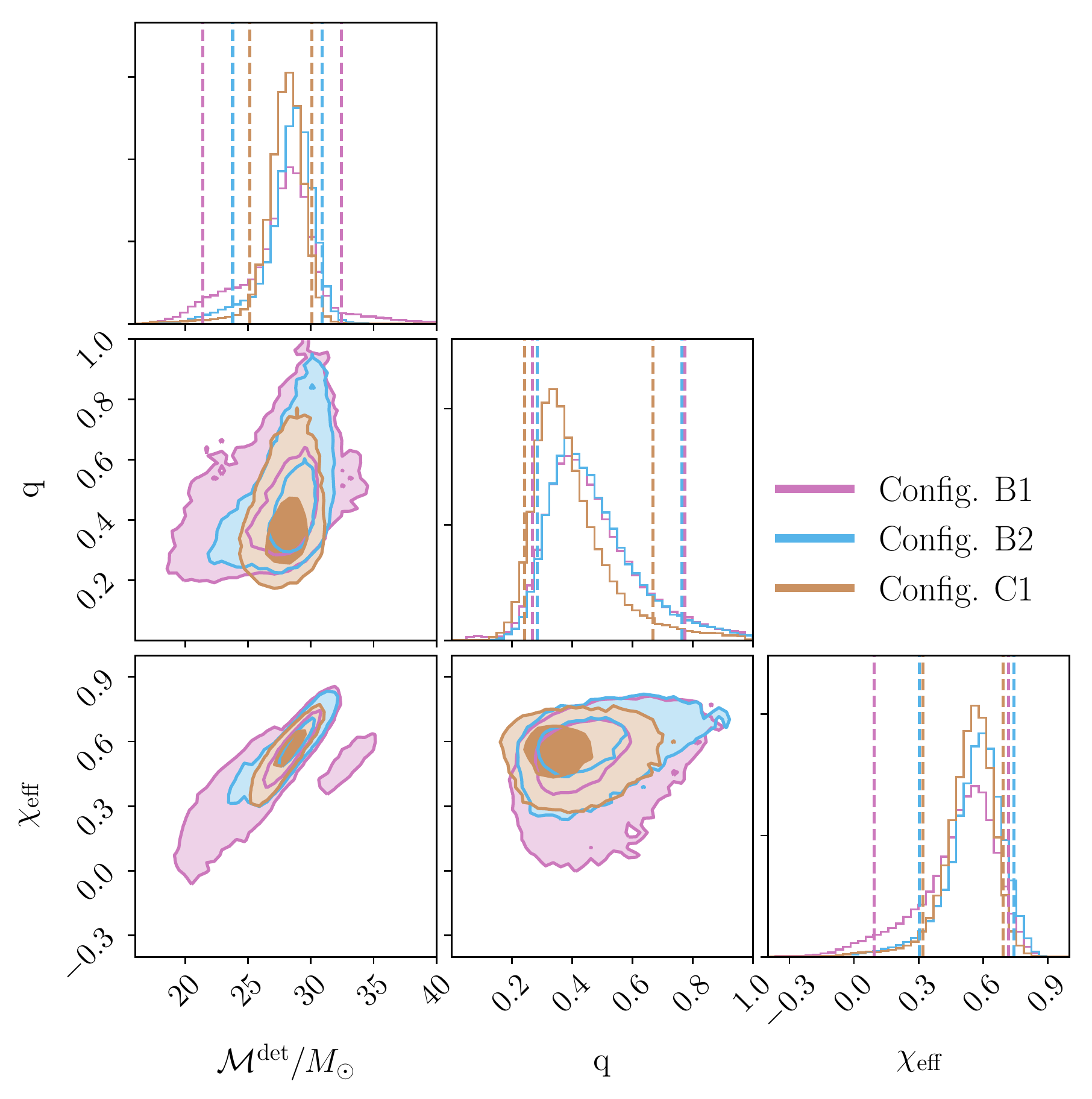}
        \caption{Effect of PSD}\label{Fig.isotropic_notched}
    \end{subfigure}
    \caption{\label{Fig.IsotropicConfigTest} Corner plot for posterior distributions for GW151216 with the isotropic prior on spins. In the left panel, green, pink and orange posteriors, respectively, use Config. B, and versions of Config. B with \textsc{pyMultinest} (B1)
    , and Config. C with the isotropic prior (C1). The blue contours in the right panel are for B2, a further modified version of B1 (data with loud lines notched out, and using the \texttt{BayesWave} continuum). Under this prior, there are residual effects of analysis choices (sampler between B and B1, treatment of lines in the data between B1 and B2, and PSD continuum between B2 and C1).
    }
\end{figure*}

The above tests were performed with the flat-in \chieff prior, and hence do not look for residual effects of the analysis methods under the alternative isotropic spin prior of Configs. A and B. To do this, we compare the results of Config. B to those of a run with a modified version of Config. C with the isotropic spin prior (henceforth Config. C1), Figure \ref{Fig.IsotropicConfigTest}. 
Similar to the comparisons in Fig.~\ref{Fig.Flatchieffcomparison}, these two runs have the same prior but differ in analysis methods (additionally, they use data segments of different length). 
We observe that (a) the differences are less pronounced than those in Sec.~\ref{Sec:GW151216}, which is consistent with our understanding that the choice of spin priors is the most significant driver of the differences in Fig.~\ref{Fig.corner151216}, and (b) unlike in Fig.~\ref{Fig.Flatchieffcomparison}, analysis methods and the data segments used make some difference here. 
The two sets of results are formally consistent with each other, but the posteriors of Config. B are broader and encompass those of Config. C1. 
In particular, the posterior on the effective spin has a fatter tail towards $ \chieff = 0$ in Config. B.

The above differences should be caused by the three remaining points of departure, i.e., the method of PSD estimation, the length of data used, and the sampler: the simplest one to vary in isolation is the sampler (analogous to Fig.~\ref{Fig.SamplerTest}). 
Toward this end, the pink contours in Fig.~\ref{Fig.SamplerTest_isotropic} show the posteriors with the first two choices fixed to those of Config. B (i.e., using the \texttt{BayesWave} PSD and a 4s segment of data), and changing the sampler from \textsc{\linf} to \textsc{pyMultinest} (henceforth, Config. B1). 
We see that contrary to the case of the `uniform in \chieff' prior (as shown in Fig.~\ref{Fig.SamplerTest}), the choice of sampler makes a small but noticeable difference here. %
The weight of the samples at $\chieff \leq 0$ is reduced to between those of Configs. B and C1.

\begin{figure}[!h]
  \centering
    \includegraphics[width=0.45\textwidth]{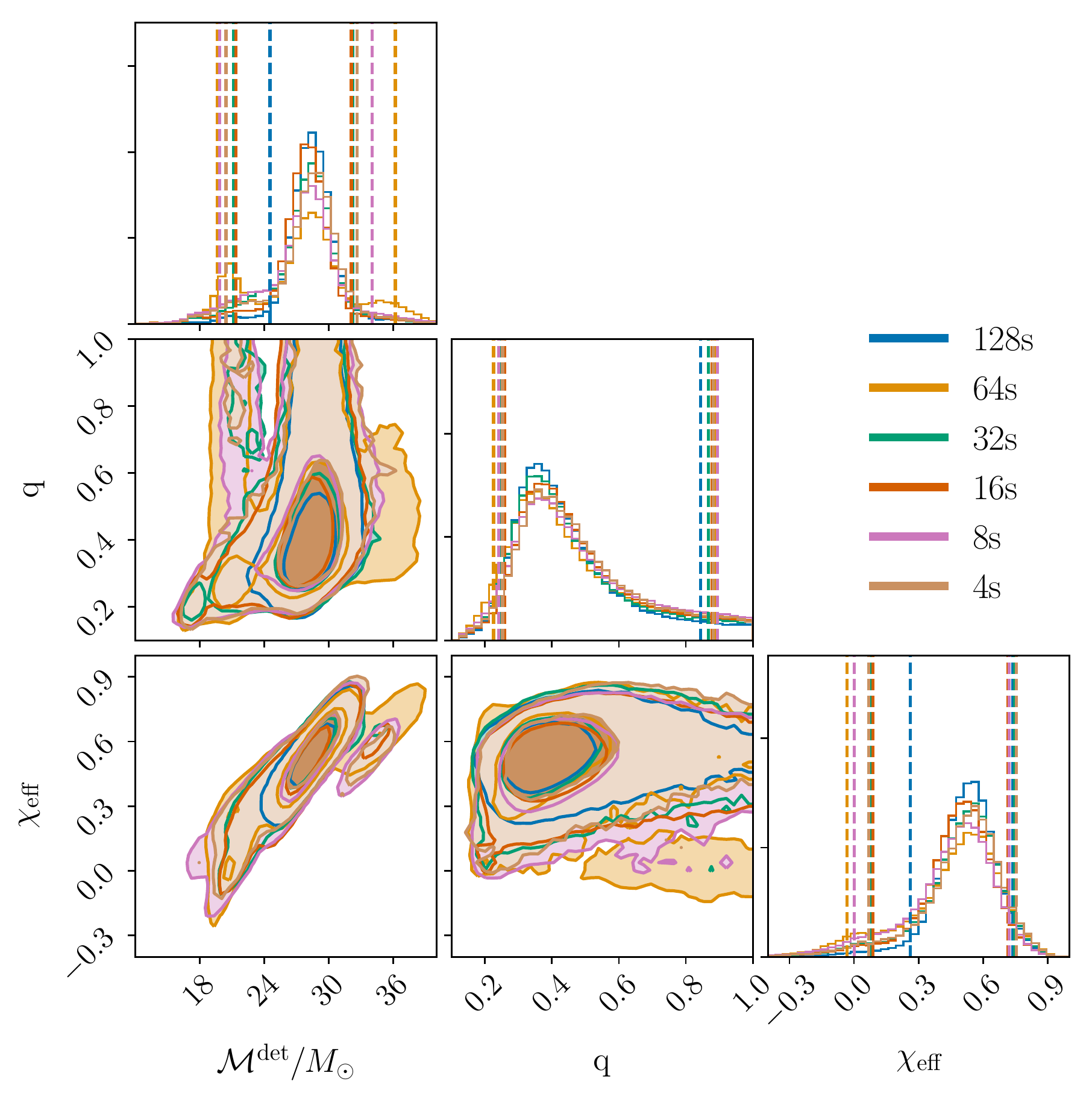}
    \caption{Corner plot for posterior distributions for GW151216, with PSD estimation method, prior and waveform choices same as Config. B. 
    The only difference is length of data segment used: 4s (brown) vs. 8s (purple) vs. 16s (red) vs. 32s (green) vs. 64s (orange) vs. 128s (blue) where all segments end 2s after the merger time (defined as peak of the absolute value of the strain amplitude at the geocenter). }
  \label{Fig.seglenTest}
\end{figure}

In Fig.~\ref{Fig.seglenTest} we show the result of varying the segment length while keeping other choices identical to Config. B. 
Here, the effect is slightly more visible, but still cannot account for the more prominent differences between the posteriors for chirp mass and effective spin seen in Fig.~\ref{Fig.corner151216} when comparing Config. B with Config. C.
A secondary mode in the spin posterior that supports zero \chieff is more visible for some segment lengths, and is most prominent for 64s and least prominent for 128s. We attribute these differences as arising from the PSD estimation. 
Different segment lengths result in slightly different PSD estimates, and these differences in the PSDs can have a visible impact on parameter estimation results for low SNR events~\cite{Huang:2018tqd}. 
We further explore assumptions used in estimating the PSDs below.

We next investigate the causes of the difference between the results under B1 and C1: these configurations differ in the method of PSD estimation, and the length of the data segment analyzed. %
The Welch method with a drift factor, used by Config. C, requires data segments much longer than 4 seconds, thus precluding a direct head-to-head comparison. A naive approach forward would be to reduce the frequency resolution of a PSD computed on longer segments (say, using the Welch method) onto a frequency grid conjugate to the 4 second segment; let us consider the validity of such an approach. The likelihood estimation in Eqs.~\eqref{Eq.Likelihood} and \eqref{Eq.Inner Product} works with the discrete Fourier transform (DFT) coefficients, $\tilde{d}(f_m)$, of the data $d$, where $f_m$ is conjugate to the 4s segment. Given the PSD, $S_{n, \rm w}(f)$, computed using the Welch method on a very fine frequency grid (over a longer segment of data), the covariance matrix of the DFT coefficients is
\begin{align}
  ~~~ & \!\!\!
  \left\langle \tilde{d} \left( f_m \right) \left[ \tilde{d} \left( f_{m^\prime} \right) \right]^\ast \right\rangle \notag \\
  & = \frac14\, e^{i\,\pi\, \left( f_{m^\prime} - f_m \right) \, \Delta t} 
  \times \notag \\
  &\int {\rm d} f\, S_{n, \rm w}(f)\, \widetilde{W} \left(f - f_m \right) \widetilde{W} \left( f - f_{m^\prime} \right), \label{eq:psddefdiscrete_simpl}
\end{align}
where $\Delta t$ is the sampling period, and $\tilde{W}$ is the Fourier transform of the window function applied to the data (typically a Tukey window). The noise PSD exhibits spectral lines that are orders of magnitude louder than the continuum, so the window function $W$ in Eq.~\eqref{eq:psddefdiscrete_simpl} induces covariances between distinct frequencies in the vicinity of the lines (i.e., between $f_m \neq f_{m^\prime}$). In such a case, the fundamental assumption in Eq.~\eqref{Eq.Likelihood}, that the frequencies can be separately analyzed when computing the likelihood, breaks down.

We avoid dealing with these complications by using a further modified version of Config. B1, in which we (a) notch out all loud spectral lines from a long segment of data, and then restrict to a 4s segment, and (b) use only the continuum of the \texttt{BayesWave} PSD to analyze this segment. This approach (henceforth Config. B2) is not perfect, but a heuristic way to contrast the effects of the data analysis choices keeping the segment length fixed. Figure \ref{Fig.isotropic_notched} contrasts the posteriors under Config. B2 to the others: we see that the posteriors in \chieff are consistent with those of Config. C, but there are residual differences in the distribution of mass-ratio $q$ (and a small bias by a fraction of a sigma in other parameters as well).

\begin{table}[tb]
\centering
\begin{tabularx}{.95\linewidth}{@{} bbbb @{}}
\toprule\toprule
 Configuration & Notes & $p(\chieff \leq 0|{\bf d})$ &  ${p(\chieff \geq 0.8|{\bf d})}$ \\ [0.5ex] 
\midrule
B & See Table \ref{Table.Configuration}  & 3.4\%  & 2.6\%\\
B1 & B + \textsc{pyMultiNest} & 2.4\% & 0.7\% \\
B2 & B1 $-$ lines & 0.3\% & 1.1\% \\
C & See Table \ref{Table.Configuration}  & 0.0\% & 52.5\% \\
C1 & C + Isotropic spin prior & 0.5\% & 0.2\%\\
 \bottomrule\bottomrule
\end{tabularx}
\caption{Posterior weight in the region $\chieff \leq 0$ for various configurations.} 
\label{Table.postweight}
\end{table} 

Table \ref{Table.postweight} reports the posterior weight in the region $\chieff \leq 0$ as well as $\chieff \geq 0.8$ for the various configurations considered in this section under the isotropic prior on spins, as well values from Config. B and C from Sec.~\ref{Sec:GW151216} for comparison. The weight of the posteriors varies (and is subject to large measurement uncertainties). $\chieff \leq 0$ contains only low-probability tails across all of the configurations considered here, while support for high spin magnitude $\chieff \geq 0.8$ is much more significant for Config. C compared to others.

\clearpage

\bibliographystyle{apsrev4-1}
\bibliography{ThisPaperBib}

\end{document}